\documentclass[iop, numberedappendix,twocolumn,twocolappendix]{emulateapj-rtx4}

\usepackage{graphicx}
\usepackage{amssymb}
\usepackage{multirow}

\newcommand{\be}{\begin{equation}}
\newcommand{\ee}{\end{equation}}

\newcommand{\chan}{{\sl Chandra }}
\newcommand{\xmm}{{\sl XMM-Newton }}

\defcitealias{HeinkeHo2010}{HH10} 	
\defcitealias{Ho2009}{HH09} 	
\defcitealias{Elshamouty2013}{E+13}

\begin{document}
 
\title{New constraints on the cooling of the Central Compact Object in Cas A}
\author{B. Posselt}
\affil{Department of Astronomy \& Astrophysics, Pennsylvania State University, 525 Davey Lab,University Park, PA 16802, USA }
\email{posselt@psu.edu}
\author{G. G. Pavlov}
\affil{Department of Astronomy \& Astrophysics, Pennsylvania State University, 525 Davey Lab,University Park, PA 16802, USA}
\author{V. Suleimanov$^1$ }
\affil{Institut f\"ur Astronomie und Astrophysik T\"ubingen, Sand 1, 72076 T\"ubingen, Germany}
\altaffiltext{1}{Kazan (Volga region ) Federal University, Kremlevskaya 18, 420008, Kazan, Russia}

\author{O. Kargaltsev}
\affil{Department of Physics, The George Washington University, Washington, DC 20052, USA}

\begin{abstract}
To examine the previously claimed fast cooling of  
the Central Compact Object (CCO) in the Cas A supernova remnant (SNR),
we analyzed two {\sl Chandra} observations of this CCO,
taken in a setup minimizing instrumental spectral distortions.
We fit the two CCO X-ray spectra from 2006 and 2012 with hydrogen and carbon 
neutron star atmosphere models. 
The temperature and flux changes in the 5.5 years between the two epochs
depend on the adopted constraints on the fitting parameters and
the uncertainties of the effective area calibrations. 
If we allow a change of the equivalent emitting region size, $R_{\rm em}$, the effective temperature remains essentially the same. 
If $R_{\rm Em}$ is held constant, the best-fit temperature change is negative,
but its statistical significance ranges from $0.8\sigma$ to $2.5\sigma$, depending on the model.
If we assume that the optical depth of the ACIS filter contaminant in 2012 was $\pm 10$\,\% different from its default calibration value, the significance of the temperature drop becomes $0.8 \sigma$ to $3.1 \sigma$, for the carbon atmospheres with constant $R_{\rm Em}$.
Thus, we do not see a statistically significant temperature drop in our data, but the involved uncertainties are too large to firmly exclude the previously reported fast cooling.
Our analysis indicate a decrease of 4\%--6\% (1.9--$2.9 \sigma$ significance) for the absorbed flux in the energy range $0.6-6$\,keV between 2006 and 2012,  most prominent in the $\approx 1.4$--1.8 keV energy range. 
It could be caused by unaccounted changes of the detector response or contributions from unresolved SNR material along the line of sight to the CCO.
\end{abstract}

\keywords{ stars: neutron --- supernovae: individual (Cassiopeia A) ---
         X-rays: stars --- X-rays: individual (CXOU J232327.8+584842)}

\section{Introduction}
\label{intro}
The physical properties of the interiors of neutron stars (NSs) are currently poorly understood. However, recent observations targeting different measurable NS properties have put some constraints on the equation of state of the NS interior;
see, e.g., \citet{Ozel2013, Hebeler2013, Steiner2010,Lattimer2007} for reviews. Assessing the cooling of a NS by studying its thermal evolution in X-rays provides a possibility to investigate the composition and structure of NSs (e.g., \citealt{Page2004}). 
An exciting result on unusually fast NS cooling was reported by \citet[HH10 hereafter]{HeinkeHo2010}.
From an analysis of several \chan  observations of the Central Compact Object (CCO) in the Cassiopeia A (Cas A) supernova remnant (SNR) they found a 4\% ($5.4\sigma$) decline of the surface temperature and a 21\% change of the flux over the time span of 10 years. 
The observed rapid cooling has been interpreted as first direct evidence for nucleon superfluidity in the core of NSs \citep{Shternin2011,Page2011}. Other models suggest a quark color superconducting  phase with a large energy gap to explain the rapid cooling \citep{Noda2013}, or cooling after an r-mode heating process \citep{Yang2011}. According to these models, a continued fast cooling can be expected for several decades. 

Different cooling model predictions can be tested by
measuring the NS surface temperature in monitoring X-ray observations of the Cas A CCO. Such measurements, however, must be extremely accurate, requiring an optimal and homogeneous instrumental setup
to minimize the systematic errors.
Recently, \citet[E+13 in the following]{Elshamouty2013} tested the temperature decline using previous observations by all {\sl Chandra} X-ray detectors in various modes (HRC-S, HRC-I, ACIS-I, ACIS-S in Faint mode and ACIS-S in Graded mode). While the results of all instruments showed indications of a temperature decrease, E+13 found it to be statistically significant only in the case of the ACIS-S  Graded mode observations where the best fit decay is $3.5\pm 0.4$\% (from 2000 to 2010).
Since the ACIS-S Graded mode observations are subject to spectral distortions because of the pile-up and charge transfer inefficiency (CTI) effects, the temperature decline needs to be further investigated with other instruments or instrument modes.
Here, we present two dedicated  observations of the Cas A CCO taken in an instrumental configuration which minimizes spectral distortions.\\ 

\begin{figure*}[t]
{\includegraphics[width=170mm]{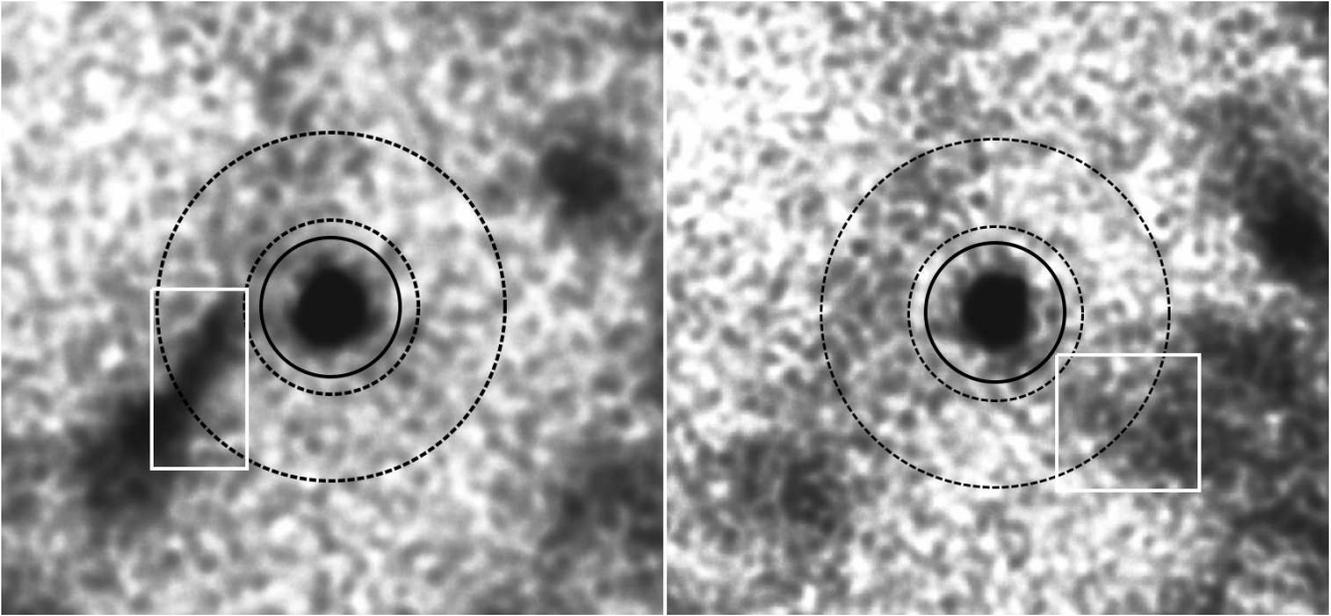}}\\ 
\caption{The left image shows our new 2012 observation, the right image shows the 2006 data. North is up, East is to the left. Marked in each image are: the inner source extraction region in black (radius of 4 ACIS pixels, $1\farcs{97}$), the annulus region for the background in black dashed lines (inner radius of 5 pixels, outer radius of 10 pixels), and the box regions excluded from the respective background regions in white.}
\label{ExtractRegions}
\end{figure*}

The Cas A SNR is very young, with an age of only $\approx 330$\,years \citep{Ashworth1980}. Its distance was determined to be $3.4^{+0.3}_{-0.1}$\,kpc \citep{Reed1995}. The SNR has been well studied from X-rays to radio wavelengths (e.g, \citealt{DeLaney2010,Helmboldt2009,Fesen2006}, and references therein).   
An X-ray point source at the SNR center was detected in the first light observations with \chan \citep{Tananbaum1999}. The properties of the point source in these observations were analyzed by \citet{Chakrabarty2001} and by \citet{Pavlov2000} who called the point source a ``Compact Central Object''. A thermal-like spectrum was found, coming from a small inferred emitting area. \citet{Pavlov2000} and \citet{Chakrabarty2001} suggested that the X-rays are emitted from hot spots on a NS, similar to some other CCOs and rotation-powered pulsars.\\ 

Since the Cas A CCO is a bright X-ray source, the first-light and many subsequent observations with the \chan ACIS instrument in full-frame mode were strongly affected by the photon pile-up effect\footnote{Two or more photons are detected as a single event; for more details see cxc.harvard.edu/ciao/ahelp/acis\_pileup.html}. Therefore, the measured spectrum of the point source was distorted.
Using an instrumental configuration designed to minimize pile-up and CTI effects, \citet{PavlovL2009} (PL09 in the following) investigated the Cas A CCO with a 62\,ks ACIS-S observation. They confirmed the thermal-like spectrum without any significant power-law component. They found that while a low-magnetic field hydrogen atmosphere model can fit the spectrum, the respective inferred emission radius is only $R_{\rm Em} \sim 4-6$\,km, smaller than the expected NS radius. PL09 interpreted this as hot spots on a NS, but did not exclude a strange quark star as a counterpart.\\    

A carbon atmosphere model for a NS with low magnetic field was shown to produce a good fit to the spectrum as well \citep[HH09 in the following]{Ho2009}. In contrast to other atmosphere models, such a carbon atmosphere implied an emission size consistent with theoretical predictions for the NS radius ($R_{\rm NS}=8-17$\,km).
\citetalias{HeinkeHo2010} used these non-magnetic carbon atmosphere models to investigate the surface temperature evolution in the Cas A CCO.\\

No X-ray or radio pulsations have been detected from the Cas A CCO so far.
A low-significance 12\,ms period has been reported by \citet{Murray2002} from \chan  HRC  observations, but it was not confirmed in later observations. From \xmm observations \citet{Mereghetti2002} derived a $3\sigma$ upper limit on the pulsed fraction of $<13$\% for $P>0.3$\,s, but PL09 argued that the limit on the background-corrected pulsed fraction is a factor of three higher. PL09 did not find  pulsations with a period $P>0.68$\,s in the \chan ACIS observations and estimated the upper limits on the pulsed fraction to be 16\% (99.9\% confidence level).
\citet{Halpern2010} obtained the lowest limit on the instrinsic pulsed fraction so far, 12\% (99\% confidence level) for $P>0.01$\,s from HRC timing observations.\\

The Cas A CCO is similar to CCOs in other young SNRs. These X-ray sources are all characterized by their thermal-like X-ray spectra, the lack of pulsar wind nebulae, and the lack of detections at wavelengths other than X-rays (e.g., \citealt{Pavlov2002a,Pavlov2004} and \citealt{deLuca2008}). 
The emitting areas inferred from blackbody fits are
substantially smaller than the expected NS surface area for all CCOs.
In contrast to the Cas CCO, however, three of the eight most secure CCOs have detected X-ray pulses with periods in the range of $0.1-0.4$\,s and pulsed fractions of 9\%, 11\% and 64\%  (e.g., \citealt{Gotthelf2013} and references therein). \citet{Gotthelf2013} and \citet{Halpern2010} measured the period derivatives for these three CCOs and inferred dipole magnetic fields of $B=(3-10) \times 10^{10}$\,G, supporting their `anti-magnetar' hypothesis for the CCOs. It is, however, not clear whether the objects are born with low magnetic fields or whether they have much higher magnetic (dipole and toroidal) fields under the crust (e.g., buried under supernova fallback material). The `hidden magnetar' or `hidden magnetic field' scenario has become a viable model for the CCOs since it can explain several observed features, e.g., the highly anisotropic surface temperature distributions (e.g., \citealt{Shabaltas2012,Vigan2012,Ho2011} and references therein).  
It remains unclear whether or not the Cas A CCO is different from other CCOs, in particular, whether or not its atmosphere has a different chemical composition.

\section{Observations and Data Reduction for the Spectral Analysis}
\label{obsred}
We report on two observations of the Cas A CCO taken in the same instrumental setup. 
The observations were done on 2006 October 19 (MJD 54027, ObsID 6690, 61.7\,ks dead-time-corrected exposure time) and on 2012 May 5 (MJD 56052, ObsID 13783, 63.4\,ks dead-time-corrected exposure time) using \chan ACIS in the Faint telemetry mode. 
In each observation the target was imaged on the ACIS-S3 chip in the 100 pixel subarray.
This reduces the frame time to 0.34\,s versus the 3.24\,s in full-frame mode. This smaller frame time  reduces the pile-up effect. The estimated pile-up fraction is only 1.6\% in comparison to $\sim 20$\,\% in the case of the full frame mode (PL09). The subarray was placed near the chip readout to reduce the CTI effect on the spectrum.

We used CIAO version 4.5 with CALDB version 4.5.5.1 for reprocessing both data sets with the latest calibrations, for producing images, extracting spectra and creating detector responses. 
Both observations were free of strong background flares. We filtered the reprocessed event files for good time intervals which have less than $5\sigma$ deviation from the overall light curve mean rate. The effective exposure times of the filtered event files were 61.7\,ks in 2006 and 63.0\,ks in 2012. 
We selected a circle with radius of 4 ACIS pixels ($1\farcs{97}$) for the source extraction region in each observation, the same as HH10 used for the 2006 observations.
For such source extraction regions, the centroid positions on the S3 chip are ($210.9, 49.3$) pixels for the 2006 observation and ($215.0, 50.8$) pixels for the 2012 observation in chip coordinates. In sky coordinates, the 2012 position appears to be slightly shifted to the north-northwest, the formal spatial separation between the 2006 and 2012 CCO centroid positions is $0\farcs{16}$. Thus, within the uncertainty of the \chan absolute astrometry, $0\farcs{4}$, the positions are consistent with each other.   
We note that \citet{DeLaney2013} estimated the proper motion of the Cas A CCO as  $390\pm 400$\,km\,s$^{-1}$. Thus, the CCO is expected to have moved by $0.0022 \pm 0.0023$\, pc between 2012 and 2006,
corresponding to a possible shift of $0\farcs{13} \pm 0\farcs{14}$. Due to the lack of reference point sources for relative astrometry we neither correct for a systematic coordinate shift nor can measure the proper motion of the CCO.\\

\begin{figure}[b]
{\includegraphics[width=70mm]{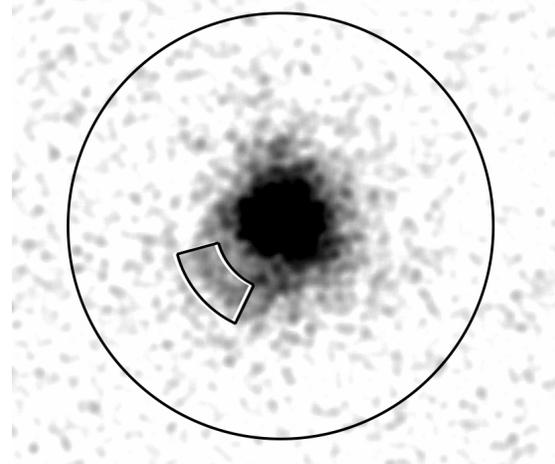}} 
\caption{The spatial distribution of events in the 2012 observation. The large circle marks the spectral extraction radius (4 ACIS pixels, $1\farcs{97}$), the other region indicates the area of the \textsl{Chandra} point spread function asymmetry (see text).}
\label{P2asymmetry}
\end{figure}

From the comparison of the 2006 and 2012 observations (Fig.~\ref{ExtractRegions}) it is obvious that the emission from the surrounding SNR filaments changed in brightness and in position with respect to the CCO. In particular, the slightly enhanced emission south-west of the CCO in 2006 has disappeared in 2012, while the filament south-east of the CCO is brighter and closer to the CCO in 2012. We remove the respective regions of enhanced emission in the background annulus, for which we chose an inner radius of 5 pixels and an outer radius of 10 pixels. \\

All spectra were binned requiring a signal to noise ratio of at least 10 in each bin.
The spectral analysis was done using XSPEC (version 12.7.1). We excluded energies higher than 5\,keV since the background dominates at these energies. We used the Tuebingen-Boulder ISM absorption model (\texttt{tbabs}) with the solar abundance table from \citet{Wilms2000}, the photoelectric cross-section table from \citet{Balu1992} together with a new He cross-section based on \citet{Yan1998}.\\

PL09 searched for extended pulsar wind nebula emission close to the CCO in the 2006 observation and found none. The 2012 observation does not reveal such extended emission either.
A region with slightly enhanced emission can be attributed to  
the asymmetry in the \textit{Chandra} point spread function\footnote{http://cxc.harvard.edu/ciao/caveats/psf\_artifact.html}. We obtained the location of the inflicted region by applying the CIAO task \texttt{make\_psf\_asymmetry\_region}, see Figure~\ref{P2asymmetry}.\\

\begin{figure}[t]
{\includegraphics[height=85mm, angle=90]{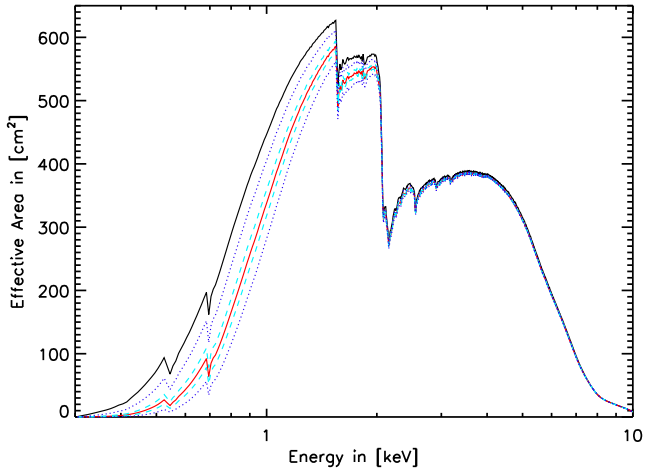}}\\ 
\caption{Effective areas for the position of the CCO target on the ACIS-S3 chip for different epochs and different contamination levels. The black and red solid lines represent the 2006 and 2012 effective areas using the current CIAO contamination model. The cyan dashed lines represent the 2012 effective area considering changes of the optical depth of the contaminant by $\pm 10$\% (in the range of model uncertainties). The blue dotted lines  indicate modifications of the contaminant's optical depth by  $\pm 30$\%.}
\label{contameff}
\end{figure}
A contaminant has been accumulating on the optical-blocking filters of the ACIS detectors. The contaminant accumulation is changing over time. The effect on the inferred spectral parameters of an X-ray source depends on the contamination model implemented in the data reduction. An error in the contamination correction may lead to a different measured flux and may offset the derived spectral parameters.  
The uncertainties of the optical depth measurements at 0.67\,keV, on which amongst others the contamination model is based, are of the order of $5$\%$-10$\% (A. Vikhlinin, {\emph{pers. comm.}}).
We changed the optical depth by $\pm 10$\% (possible changes) and $\pm 30$\% (less likely changes) for trial spectra derived from the 2012 data  to investigate the effects on our spectral analysis results. The resulting effective areas are shown in Figure~\ref{contameff}. For details on the contamination model and our modifications we refer to the Appendix ~\ref{contamtest1}.

\section{Results}
\subsection{Spectral Analysis}
PL09 presented a detailed spectral analysis for the 2006 observation. We concentrate in the following on atmosphere model fits but note that we checked the simpler model fits (e.g., blackbody, power law or a combinations of those) for consistency as well.    
\subsubsection{Carbon atmosphere models}
\label{resultscarbon}
The main goal of our new observations is to examine the temperature drop reported by HH10 who fitted multi-epoch X-ray data with the carbon atmosphere models by HH09. For our XSPEC fits, we employed new carbon atmosphere models\footnote{The carbon atmosphere models by HH09 are not public.}, which are briefly summarized in the Appendix~\ref{carbmodels} and will be presented in detail in another publication (Suleimanov  et al., { \emph{subm. to ApJS}}). 

\begin{figure}[t]
{\includegraphics[width=85mm, bb=15 15 218 170,clip]{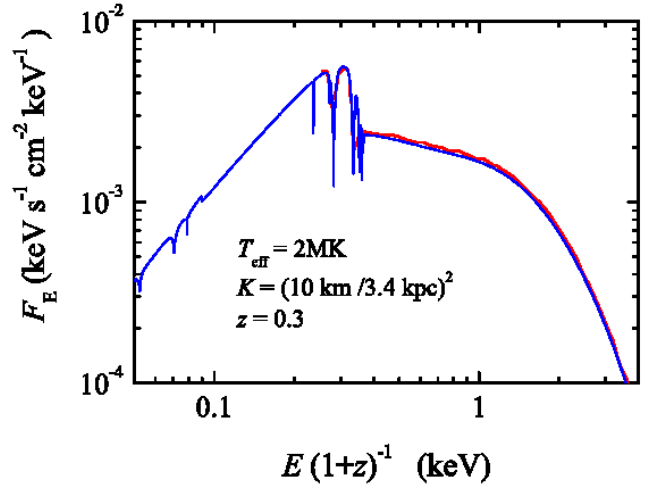}} 
\caption{The carbon model atmosphere spectra developed by HH09 (their Figure 2) and by us are plotted for comparison. HH09 (red line) assumed $M=1.4 $\,M$_\odot$, $R_{\rm NS}=10$\,km, corresponding to a gravitational redshift $z = 0.3$ and $\log g=14.38$, where $g$ is the gravitational acceleration on the surface of the NS defined by Equation~\ref{eq:gdef}, and $z$ is defined by Equation~\ref{eq:zdef} in the Appendix~\ref{carbmodels}. Our model for $z = 0.3, \log g=14.3$ is shown with the blue line.
\label{carbonmodel}}
\end{figure}

Here, we note only that these atmosphere models are very similar to those  by HH09 as demonstrated in Figure~\ref{carbonmodel}. 
This is particularly true for the observable energy range -- in the case of the Cas A CCO there is nearly no detected X-ray flux for energies below 1\,keV. Therefore, the fit parameters obtained by fitting the spectra with our carbon atmosphere models are expected to be consistent with those that would have been obtained with the models by HH09.\\

\begin{figure}[b]
{\includegraphics[height=85mm, bb=75 15 565 702, angle=270, clip]{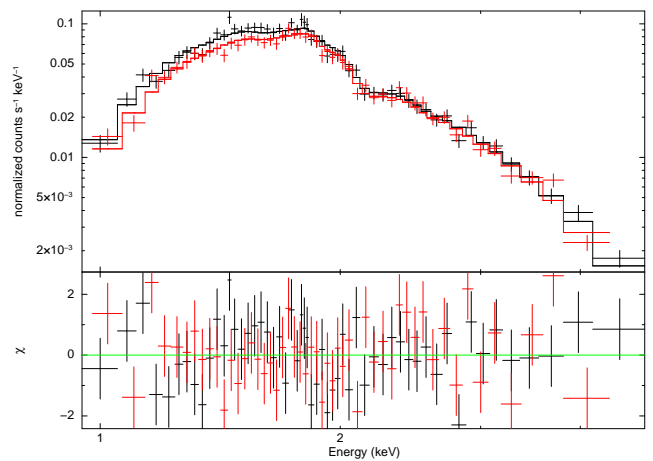}} 
\caption{The data and our fit to the ($\log g=14.45$, $z=0.375$) carbon atmosphere model in the case of different $N_{\rm H}$ values for the first epoch (black) and the second epoch (red). For this fit, it is assumed that the whole NS surface is emitting in X-rays and that the distance is 3.4\,kpc. The count rates at energies $<2$\,keV are lower in the second epoch because of the decreasing effective area (mostly due to the increasing ACIS filter contamination -- see Figure~\ref{contameff}).
We use the default ACIS filter contamination in this fit. The lower panel shows the fit residuals in units of sigmas. 
\label{carbondatres}}
\end{figure}

Our carbon atmosphere models were calculated for nine different surface gravitational accelerations, $g$, ranging from $\log g=13.7$ to $\log g = 14.9$ with a sampling of 0.15. 
As demonstrated in the Appendix~\ref{carbmodels}, it does not matter quantitatively which ($\log g$, $z$) pair one chooses to investigate a possible temperature evolution -- as long as this pair is located within our $1\sigma$ mass-radius contour.
We will give in the following all our fit results for ($\log g=14.45$, $z=0.375$)\footnote{This corresponds to a NS with $M_{\rm NS}=1.647$\,M$_{\odot}$ and  $R_{\rm NS}=10.33$\,km.}, which is the nearest point of our ($\log g$, $z$) grid to the ($\log g=14.4534$, $z=0.377$) used by HH10.\\

We considered two cases for the X-ray absorption along the line of sight.
In one set of models we required the absorbing $N_{\rm H}$ to be the same in the two observing epochs, in the other set $N_{\rm H}$ was allowed to be different. 
Different $N_{\rm H}$ could indicate a changing ISM density along the line of sight, but it could also account for matter distribution changes in the SNR itself. The possibility of a $N_{\rm H}$  change is supported by several observations. The $N_{\rm H}$ distribution map by \citet{Hwang2012} (their Figure 1) shows asymmetries with localized $N_{\rm H}$ enhancements of $\approx 50$\%. In the immediate surrounding of the CCO, there are differences of $\triangle N_{\rm H} \approx 2 \times 10^{21}$\,cm$^{-2}$ on a scale of $5\arcsec$. Regarding the SNR itself, asymmetric matter distributions in shocked (outer SNR) and unshocked (central part filled with cold dust) regions are reported in great detail, e.g., by \citet{DeLaney2010, Barlow2010, Rest2011}. 
The CCO environment is also highly dynamic -- the SNR as a whole appears to move to the north with a plane of the sky velocity of about 700\,km\,s$^{-1}$ \citep{Hwang2012}, while the CCO's (highly uncertain) proper motion corresponds to a velocity of around 400\,km\,s$^{-1}$ in the south-east direction (e.g., \citealt{DeLaney2013}, \citealt{Fesen2006}). 
It is reasonable to allow for the possibility of a changing effective absorption by the SNR material in the line of sight to the CCO given the overall dynamic and inhomogeneous environment.
Similarly, the ISM around the Cas A SNR is known to be very inhomogeneous too, it is assembled in fractal structures which indicate turbulent motions \citep{Kim2008, Fesen2011}. Therefore, we also cannot exclude $N_{\rm H}$ changes from turbulent ISM clumps. 
Different $N_{\rm H}$ values could also have their origin in an imperfect contamination correction of the ACIS instrument.\\

We note that, strictly speaking, the extinction models, such as \texttt{tbabs}, are not directly applicable to the absorption by the hot SNR material because the element abundances and temperatures are very different to those in the ISM. We, however, have to use the ISM extinction models as approximate description because of the lack of models for intra-SNR extinction.\\

\begin{deluxetable*}{lcccccccc}[t]
\tablecaption{Fit results for the carbon atmosphere models with $\log g=14.45$ and $z=0.375$  \label{table:carbonfits}}
\tablewidth{0pt}
\tablehead{
\colhead{Data} & \colhead{Norm} & \colhead{$N_{\rm H}$} & \colhead{$T_{\rm eff}$} & \colhead{$\triangle T_{\rm eff}$} & \colhead{$F^{\rm{abs}}_{-13}$} & \colhead{$F^{\rm{unabs}}_{-12}$}  & \colhead{$L_{\rm bol}^{\infty}$} & \colhead{$\chi^2_{\nu}$/d.o.f.} \\
\colhead{ }  & \colhead{} & \colhead{$[10^{22}$\,cm$^{-2}]$} & \colhead{[$10^4$ K]} & \colhead{[$10^4$ K]} & \colhead{} & \colhead{} & \colhead{$[10^{33}$\,erg\,s$^{-1}]$} & \colhead{}   
}
\startdata
\\
\multicolumn{9}{c}{{{Normalizations fixed }}}\\
\\
\tableline
\\
O1  & 923 & $2.27 \pm 0.05$ & $201.4^{+1.1}_{-1.0}$ & $\cdots$ & $7.37 \pm 0.17$  & $2.93 \pm 0.10$ & $6.6 \pm 0.1$ & 1.062/103\\ [0.7ex]
O2  & 923 & $=N_H$(O1) & $199.7 \pm 1.1$ & $-1.8 \pm 1.2$ & $6.96 \pm 0.17$  & $2.79^{+0.10}_{-0.09} $ & $6.4 \pm 0.1$ & 1.062/103 \\[0.7ex]
 &  &  &  & ($-0.9$\% $T_{\rm O1}$) & ($-5.6$\% $F^{\rm{abs}}_{\rm O1}$) &  ($-4.8$\% $F^{\rm{unabs}}_{\rm O1}$)  \\
\\
O1  & 923 &  $2.25 \pm 0.06$ & $201.1 \pm {1.2}$ & $\cdots$ & $7.34 \pm 0.18$  & $2.90^{+0.12}_{-0.11} $ &  $6.6 \pm 0.2$ & 1.063/102\\[0.7ex]
O2  & 923 &  $2.30 \pm 0.07$ & $200.1^{+1.3}_{-1.4} $ & $-0.9\pm 1.8$ & $6.99 \pm 0.18$ & $2.83\pm 0.12$ &  $6.4 \pm 0.2$ & 1.063/102 \\[0.7ex]
 &  &   &  & ($-0.5$\% $T_{\rm O1}$) & ($-4.8$\% $F^{\rm{abs}}_{\rm O1}$)  & ($-2.4$\% $F^{\rm{unabs}}_{\rm O1}$)\\
\\
\tableline
\\
\multicolumn{9}{c}{{{Normalizations tied, but free }}}\\
\\
\tableline
\\
O1  & $800^{+270}_{-200}$ & $2.23 \pm 0.10$ & $205.9^{+9.5}_{-9.0}$ & $\cdots$ & $7.35 \pm 0.20$  & $2.82^{+0.11}_{-0.10}$ &  $6.3^{+0.7}_{-0.6}$ & 1.066/102\\[0.7ex]
O2  & $800^{+270}_{-200}$ & $=N_H$(O1) & $204.0^{+9.4}_{-8.9}$ & $-1.8^{+1.2}_{-1.3}$ & $7.01 \pm 0.17$  & $2.66^{+0.14}_{-0.13}$  & $6.0^{+0.7}_{-0.6} $ &  1.066/102    \\[0.7ex]
 &  &  &  & ($-0.9$\% $T_{\rm O1}$) & ($-4.6$\% $F^{\rm{abs}}_{\rm O1}$) &  ($-5.7$\% $F^{\rm{unabs}}_{\rm O1}$)  \\
\\
O1  & $810^{+280}_{-200}$ &  $2.21^{+0.11}_{-0.10}$ & $205.3^{+9.4}_{-9.3}$ & $\cdots$ & $7.37^{+0.18}_{-0.17}$  & $2.83 \pm 0.11 $ &  $6.2^{+0.8}_{-0.6} $ &1.068/101\\[0.7ex]
O2  & $810^{+280}_{-200}$ &  $2.26 \pm 0.12$ & $204.2^{+9.3}_{-9.0} $ & $-1.1\pm 1.9$ & $7.02 \pm 0.18$ & $2.75 \pm 0.13$ &  $6.1^{+0.7}_{-0.6} $ &1.068/101 \\[0.7ex]
 & &   &  & ($-0.5$\% $T_{\rm O1}$)& ($-4.7$\% $F^{\rm{abs}}_{\rm O1}$)  & ($-2.8$\% $F^{\rm{unabs}}_{\rm O1}$)\\
\\
\tableline
\\
\multicolumn{9}{c}{{{Normalizations free and untied}}}\\
\\
\tableline
\\
O1  & $880^{+440}_{-290}$ &  $2.23^{+0.26}_{-0.13}$ & $202.5^{+12.9}_{-12.1}$ & $\cdots$ & $7.35 \pm 0.20$  & $2.88^{+0.27}_{-0.24} $ & $6.5^{+1.1}_{-0.8} $ &1.076/100\\[0.7ex]
O2  & $740^{+390}_{-250}$ &  $2.23 \pm 0.16$ & $207.2^{+13.9}_{-13.1} $ & $+5 \pm 19 $ & $7.04 \pm 0.20$ & $2.69^{+0.28}_{-0.25}$ &$5.9^{+1.0}_{-0.8} $ & 1.076/100 \\[0.7ex]
 & &   &  & ($+2.3$\% $T_{\rm O1}$) & ($-4.3$\% $F^{\rm{abs}}_{\rm O1}$) & ($-6.6$\% $F^{\rm{unabs}}_{\rm O1}$)\\
\\
\tableline
\enddata

\tablecomments{The fits were done simultanously for O1 and O2 (first and second observation epoch, respectively). The normalization is defined as $\mathcal{N}=R^2_{\rm NS} / d_{\rm 10kpc}^2$, where  $R_{\rm NS}$ is the NS radius in km, and $d_{\rm 10kpc}$ is the distance in 10\,kpc. $\mathcal{N}=923$ corresponds to the distance of 3.4\,kpc assuming emission from the whole uniformly heated surface of a neutron star with $\log g=14.45$ and $z=0.375$. 
Fluxes are given for the energy range of 0.6-6\,keV. $F^{\rm{abs}}_{-13}$ is the absorbed flux in units of $10^{-13}$\,erg\,cm$^{-2}$\,s$^{-1}$, while $F^{\rm{unabs}}_{-12}$ is the unabsorbed flux in units of $10^{-12}$\,erg\,cm$^{-2}$\,s$^{-1}$. All errors indicate the 90\% confidence level for one parameter of interest.
The bolometric luminosity at inifinity is calculated as $L^{\infty}_{\rm bol}=4 \pi \sigma {R^{\infty}_{\rm Em}}^2 {T^{\infty}_{\rm eff}}^4= 4 \pi \sigma 10^{10} \mathcal{N} d_{\rm 10kpc}^2 {T_{\rm eff}}^4 (1+z)^{-2}$\,erg\,s$^{-1}$. 
The temperature differences were calculated using the temperature values before rounding. The temperature uncertainties were determined from the 90\% confidence contours for one parameter of interest, which are a factor 0.77 the contour levels for two parameters of interest shown in, e.g., Figures~\ref{carbonT1T2big} and \ref{carbonT1T2zoom}.
Similarly, the luminosity uncertainties were determined from the 90\% confidence contours in temperature and normalization, e.g., in Figure~\ref{carbonNormfreeT}, if the normalizations were fit parameters.  
} 
\end{deluxetable*}

\begin{figure}[t]
{\includegraphics[width=85mm, bb=25 12 570 545, clip]{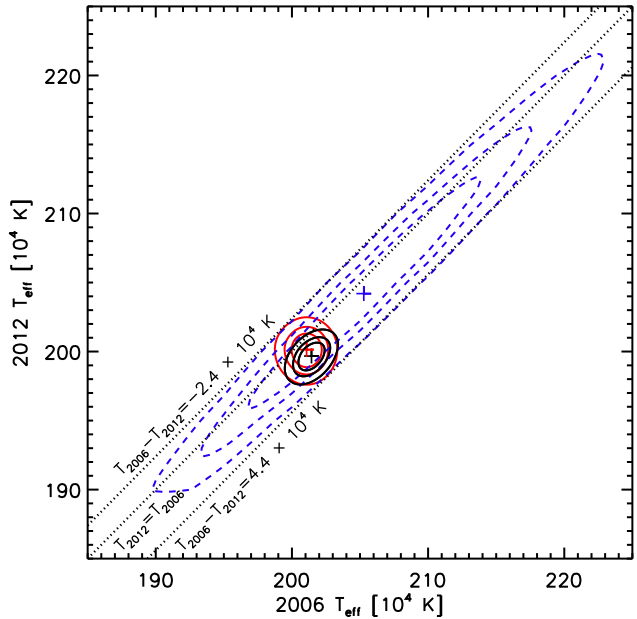}} 
\caption{Temperature confidence contours (68\,\%, 90\,\%, 99\,\%) for the fit to the ($\log g=14.45$, $z=0.375$) carbon atmosphere model for the default ACIS filter contamination.  Note that in this and all following contour plots we mark the contour levels for two parameters of interest. The black solid contours mark the fit where the $N_{\rm H}$ is set to be the same for both epochs, the red dashed contours mark the fit where the $N_{\rm H}$ is allowed to be different (see Figure~\ref{carbonT1T2zoom} for a zoom-in version). In both cases, the distance is fixed at 3.4\,kpc and the whole NS surface is assumed to emit in X-rays. 
In contrast to that, the blue dashed contours mark the fit where the normalization is a fit parameter, although it is tied between the epochs. $N_{\rm H}$ is allowed to be different in the two epochs for this fit. See Table~\ref{table:carbonfits} and text for more details on the individual fits. The dotted black lines mark lines of constant temperature difference as indicated.
\label{carbonT1T2big}}
\end{figure}

\begin{figure}[b]
{\includegraphics[width=85mm, bb=25 12 570 545, clip]{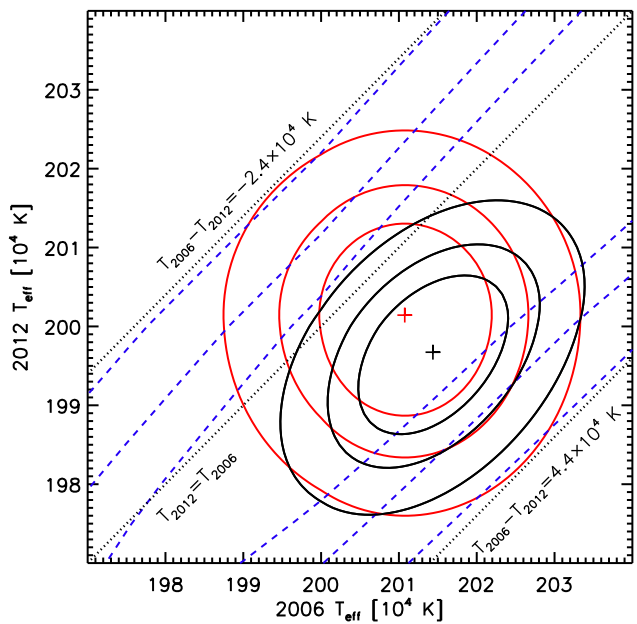}} 
\caption{Enlarged view on the central part of Figure~\ref{carbonT1T2big} employing the same layout. 
\label{carbonT1T2zoom}}
\end{figure}

\begin{figure}[t]
{\includegraphics[width=85mm, bb=25 12 570 540, clip]{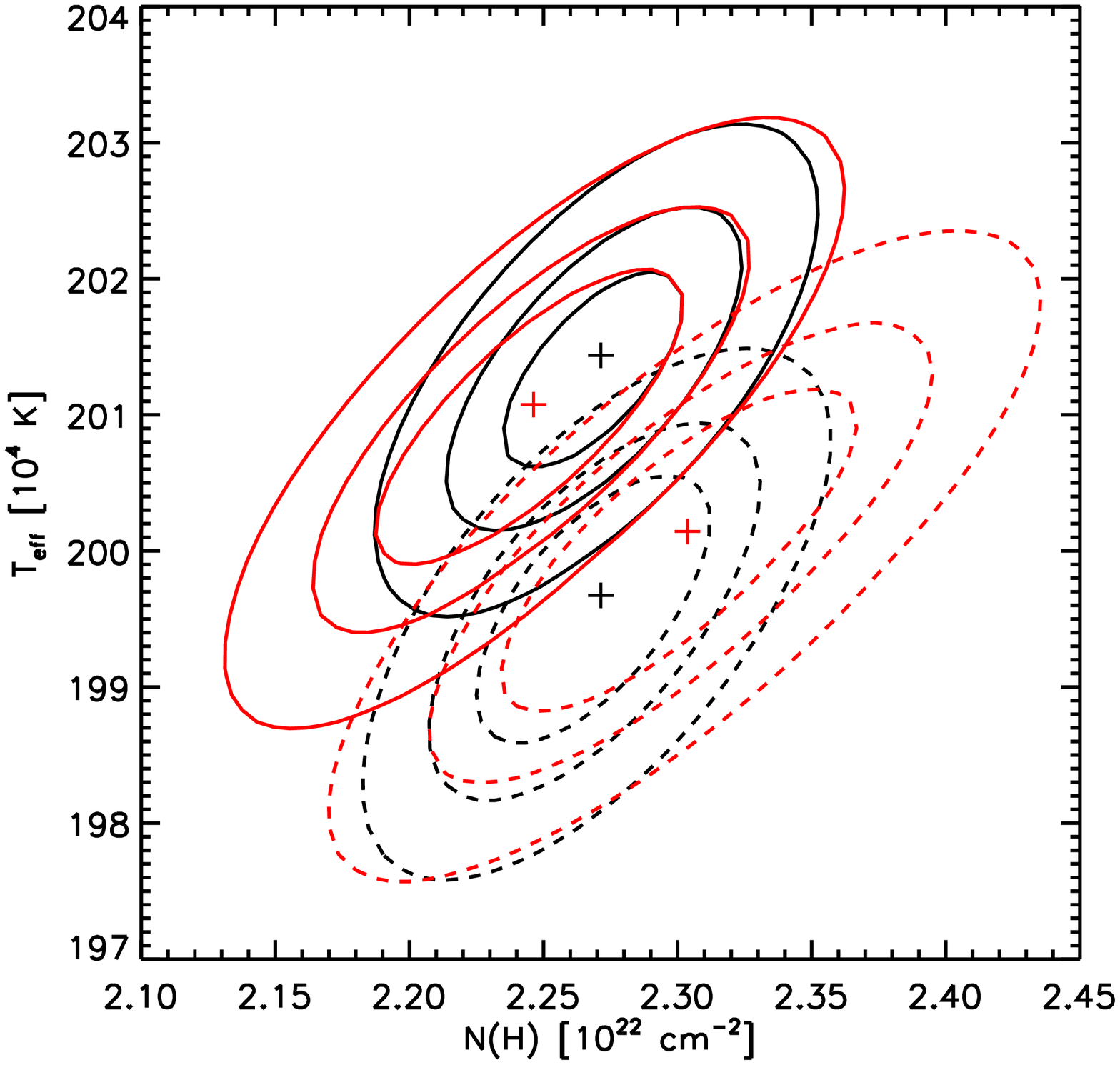}} 
\caption{Temperature versus $N_{\rm H}$ confidence contours (68\,\%, 90\,\%, 99\,\%) for the fit to the ($\log g=14.45$, $z=0.375$) carbon atmosphere model.
For these fits, it is assumed that the whole NS surface is emitting in X-rays and that the distance is 3.4\,kpc. The black contours mark the fit where $N_{\rm H}$ is set to be the same for both epochs, the red contours mark the fit where the $N_{\rm H}$ is allowed to be different. The solid line contours indicate the 2006 values while the dashed line contours indicate the 2012 values. For more details see  Table~\ref{table:carbonfits} and text.
\label{carbonNHT}}
\end{figure}

The fit results for the different sets of models are listed in Table~\ref{table:carbonfits}. In Figure~\ref{carbondatres}, we show the fit and its residuals for the case of different $N_{\rm H}$ in the two epochs, assuming a fixed distance of 3.4\,kpc and X-ray emission from the whole uniformly heated NS surface. 
In general, our carbon atmosphere models describe the data well. The $\chi^2_{\nu}$ values of our best fits to the carbon atmosphere models are slightly smaller than those of the best hydrogen atmosphere model fits (see Section~\ref{resultsnsa}). Allowing $N_{\rm H}$ to vary between the two epochs results in slightly different $N_{\rm H}$ values, which are, however, still consistent with each other within errors.\\

The confidence contours of the 2006 and 2012 temperatures are shown in Figures~\ref{carbonT1T2big} and \ref{carbonT1T2zoom}, the latter being a zoom-in version of the central part of the former. The models represented in these figures consider three cases: tied and untied $N_{\rm H}$ values for a fixed normalization (which assumes that the whole NS is emitting at $d=3.4$\,kpc) and untied $N_{\rm H}$ values for free but tied normalizations (allowing an uncertainty in either the NS emission area and/or the NS distance).
In Table~\ref{table:carbonfits} we also list the case of free and untied normalizations (allowing the emission area to be different in the two epochs). 
It is evident from the fit results in Table~\ref{table:carbonfits} and Figure~\ref{carbonT1T2big} that the errors of the individual temperatures are significantly larger (about a factor 9) if the carbon atmosphere normalization (i.e., distance or emitting area) is a free fit parameter. 
Yet, Figure~\ref{carbonT1T2big} also shows that, due to the strong correlations of the fit parameters, the uncertainty in the {\emph{ temperature difference}}
is virtually the same -- regardless of whether the normalization is a fit parameter or not.
Therefore, we can restrict our discussion of the temperature change in the following to the cases of fixed normalizations, although the absolute temperatures are in fact much more uncertain.
Note that in the case of free and untied normalizations, the best fit actually indicates an (insignificant) temperature \emph{in}crease from 2006 to 2012.
Figure~\ref{carbonT1T2zoom} provides an enlarged view on the comparison of the two cases with fixed normalizations -- tied and untied $N_{\rm H}$ in the two epochs.
While the temperatures of the two epochs are consistent with each other within the 68\% confidence contour in the case of varying $N_{\rm H}$, they are consistent with each other only within the 99\% confidence contour if $N_{\rm H}$ is the same for both epochs. Thus, in general, although the best-fit temperature in 2012 is found to be slightly lower than in 2006, the temperature difference is less than 1\%, and its significance is $<3\sigma$.\\

\begin{figure}[b]
{\includegraphics[width=85mm, bb=25 12 570 540, clip]{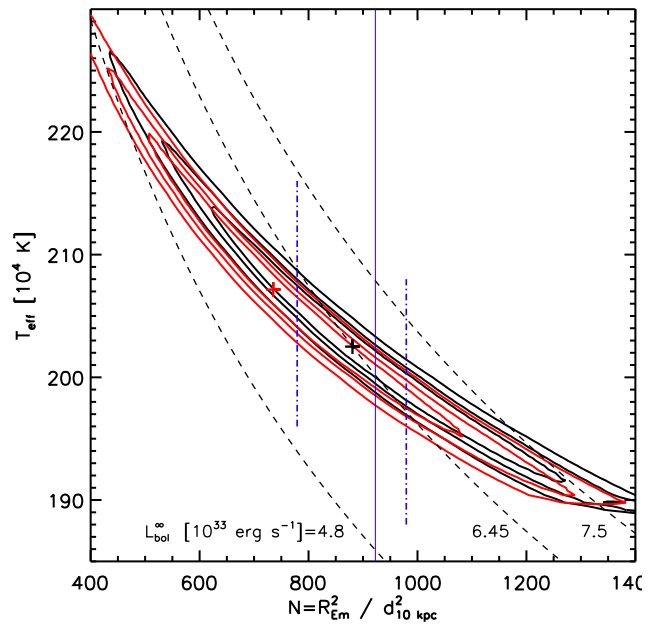}} 
\caption{Temperature versus normalization confidence contours (68\,\%, 90\,\%, 99\,\%) for the fit to the ($\log g=14.45$, $z=0.375$) carbon atmosphere model with different free normalizations, $\mathcal{N}=R^2_{\rm Em} d^{-2}_{\rm 10 kpc}$, where $R_{\rm Em}$ (in km) is the radius of an equivalent sphere with the same emission area, and $d_{\rm 10 kpc}$ is the distance in units of 10\,kpc. If the whole uniformly heated NS surface is emitting at $d=3.4$\,kpc, the expected normalization is 923 (marked with the vertical blue line). The distance uncertainty of ($+0.3$, $-0.1$)\,kpc translates into a normalization range for the uniformly emitting NS (marked by the dotted-dashed blue lines).
 The 2006 and 2012 confidence contours are plotted in black and red, respectively. The dashed black lines are lines  of constant bolometric luminosity at infinity $L^{\infty}_{\rm bol}$ for a distance of 3.4\,kpc. 
\label{carbonNormfreeT}}
\end{figure}

Figure~\ref{carbonNHT} shows confidence contours for temperatures and $N_{\rm H}$ in the case of fixed normalizations. The derived temperatures are strongly correlated\footnote{Notice that this is a positive correlation because of the fixed normalizations -- in contrast to the negative correlation for the hydrogen atmosphere fits (Figure~\ref{fig:nsaNHT}) where the hydrogen atmosphere model normalizations are free parameters.} with the equivalent hydrogen column density $N_{\rm H}$. 
Although the best-fit values are different in the two epochs, they are consistent with each other within the 90\% confidence contour for the two  model setups -- the one where $N_{\rm H}$ is set to be the same for both epochs, and the other one where $N_{\rm H}$ is allowed to be different.\\ 

The temperatures are also strongly correlated with the normalization(s) if tied or free normalizations are fit parameters. The normalization is defined as $\mathcal{N}=R^2_{\rm NS} d^{-2}_{\rm 10 kpc}$, where $d_{\rm 10 kpc}$ is the distance in units of 10\,kpc, and $R_{\rm NS}$ is in km. If only a part of the NS surface is emitting, one can formally define an equivalent sphere with radius $R_{\rm Em}$ such that $\mathcal{N}=R^2_{\rm Em} d^{-2}_{\rm 10 kpc}$.
Fitting for the normalizations allows one to discuss the following effects. First, the distance has an uncertainty,  $d=3.4^{+0.3}_{-0.1}$\,kpc \citep{Reed1995}. A distance different from 3.4\,kpc would equally influence the normalizations at both epochs. Second, the effective emission area might change if only a part of the NS surface is emitting in X-rays. This effect would lead to different normalizations in the two epochs. 
Table~\ref{table:carbonfits} lists our fit results for all these cases. Figure~\ref{carbonNormfreeT} shows confidence contours for temperatures and normalizations in the case of a fit where both normalizations are free fit parameters. The normalizations are consistent with the value expected for the case of the whole NS surface emitting at 3.4\,kpc. The uncertainties of independent normalizations overlap as well, although the actual best-fit effective emission areas are slightly different in the two epochs.\\ 

\begin{figure}[b]
{\includegraphics[width=85mm, bb=25 12 570 540, clip]{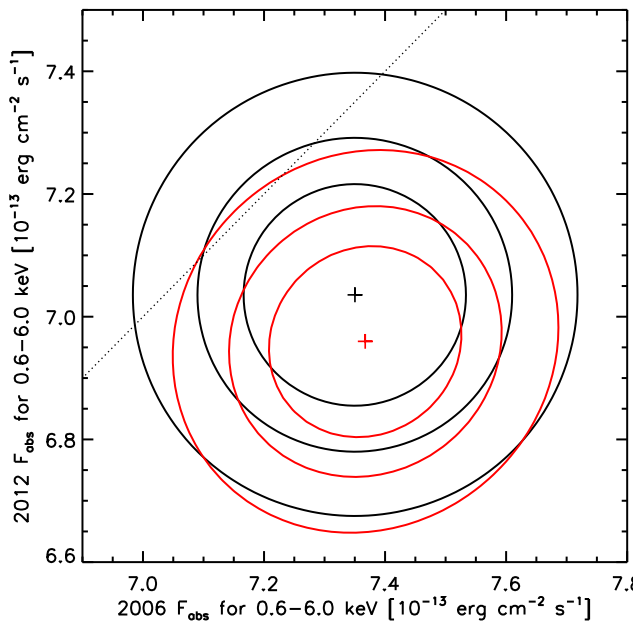}}
\caption{The confidence contours (68\,\%, 90\,\%, 99\,\%) of the absorbed fluxes derived from the fit to the ($\log g=14.45$, $z=0.375$) carbon atmosphere model using the \texttt{cflux} model component in XSPEC. Black contours represent the results of the model with free normalizations and free $N_{\rm H}$, red contours represent the results of the model with fixed normalizations and the same $N_{\rm H}$ in both epochs.
The energy range for these absorbed fluxes is 0.6\,keV to 6.0\,keV, the same energy range used for the flux values in Table~\ref{table:carbonfits}.  
The straight dotted line is the line of equal fluxes in the two epochs. 
\label{fig:cfluxbig}}
\end{figure}

We used the \texttt{cflux} model to estimate the absorbed and unabsorbed fluxes and their confidence levels in the energy range 0.6\,keV to 6\,keV, see Table~\ref{table:carbonfits}.
The best-fit 2012 fluxes are lower than the 2006 fluxes. The best-fit absorbed flux changes range from $4$\% to 6\%, depending on whether parameters ($N_{\rm H}$, $\mathcal{N}$) have been tied or not. The range for the unabsorbed flux changes is 2\% to 7\%.
The unabsorbed fluxes at the two epochs are consistent with each other at the 90\% confidence level regardless of the model setup ($N_{\rm H}$ the same or different, $\mathcal{N}$ free, tied or fixed).\\ 

Figure~\ref{fig:cfluxbig} shows the confidence contours of the absorbed fluxes for free untied $\mathcal{N}$ and free $N_{\rm H}$. The difference of the absorbed flux is $\triangle F^{\rm{abs}}_{-13}=-0.31 \pm 0.28$ (90\% confidence level for one parameter of interest). 
If the parameters are tied, the confidence contours look similar, but the flux difference can be more significant (Figure~\ref{fig:cfluxbig}). For instance, if the normalizations are fixed and  $N_{\rm H}$ is set to be the same, the difference of the absorbed flux is $\triangle F^{\rm{abs}}_{-13}=-0.41 \pm 0.23$. For all considered models, however, the differences of the absorbed fluxes are still below the $3\sigma$ significance level.\\

\begin{figure}[t]
{\includegraphics[width=79mm, bb=25 12 570 540, clip]{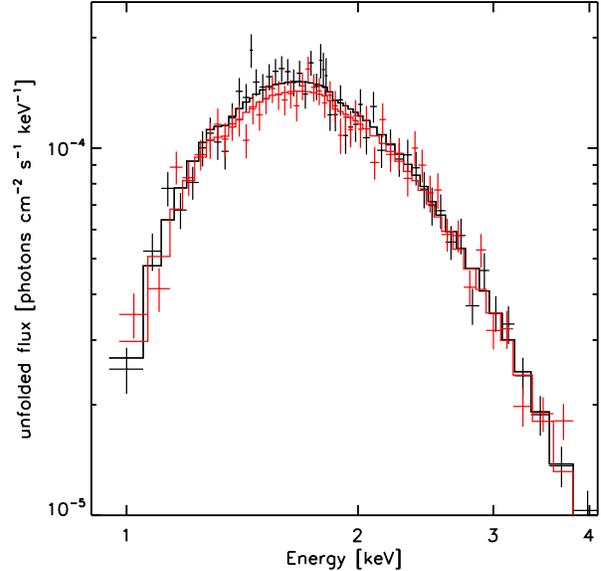}} 
\caption{The unfolded flux using the fit to the ($\log g=14.45$, $z=0.375$) carbon atmosphere model with free $\mathcal{N}$ and free $N_{\rm H}$ for the first epoch (black) and second epoch (red). The energy range 1.1-2\,keV is responsible for the apparent flux decrease.
\label{ufflux}}
\end{figure}

To investigate the origin of the apparently lower fluxes in the second epoch, we obtained the `unfolded' flux spectrum (Figure~\ref{ufflux}) using XSPEC and our best-fit carbon atmosphere model with free $\mathcal{N}$ and free $N_{\rm H}$ (Table~\ref{table:carbonfits}). 
As a cautionary note we emphasize that the data points in the unfolded flux spectra depend on the applied model.
The photon fluxes of the 2012 observation are lower than the 2006 photon fluxes in the energy range 1.1\,keV to 2.0\,keV, most prominently for $\approx 1.4-1.8$\,keV, and more so for the data points than for the model values.
Applying the \texttt{cflux} model, the differences of the absorbed fluxes in these energy ranges are $\triangle F^{\rm{abs}}_{-13}=-0.17 \pm 0.10$ for $1.1-2.0$\,keV and $\triangle F^{\rm{abs}}_{-13}=-0.09 \pm 0.06$ for $1.4-1.8$\,keV (both 90\% confidence levels). Thus, the significance for the flux drop increases from $1.8\sigma$ to $2.8\sigma$ for the carbon atmosphere model with free $\mathcal{N}$ and free $N_{\rm H}$ if a smaller energy range is considered.\\

This deficit is not due to the decreasing effective area (assuming the default calibration is correct), contrary to the deficit in the 2012 count rate spectrum (Fig~\ref{carbondatres}). 
The strongest hint of a flux decrease in the data points is seen in the energy range of $\approx 1.4-1.8$\,keV. At the same time we see no flux decrease at energies $\gtrsim 2$\,keV, which suggests that the overall apparent flux decrease is not caused by a changing NS surface temperature. Since the unfolded spectra are model dependent, we also checked the result for the model with the largest absorbed flux change and a temperature drop (fixed $\mathcal{N}$, tied $N_{\rm H}$; see Table~\ref{table:carbonfits}). The residuals with respect to the model persist nearly unchanged for this model. Therefore, we conclude that we see flux deviations from the model in the energy range of $\approx 1.4-1.8$\,keV for all the models considered in Table~\ref{table:carbonfits}, and this effect cannot be explained by a temperature change.
Interestingly, there is an excess of the data points with respect to the 2006 model spectrum and a lack of such excess in the 2012 data (Figure~\ref{ufflux}). This may hint at an inconsistency in the default ACIS responses or a varying background contribution unaccounted for in our analysis.\\

\subsubsection{Hydrogen atmosphere models}
\label{resultsnsa}

\begin{figure}[b]
{\includegraphics[height=85mm, bb=75 15 565 702, angle=270, clip]{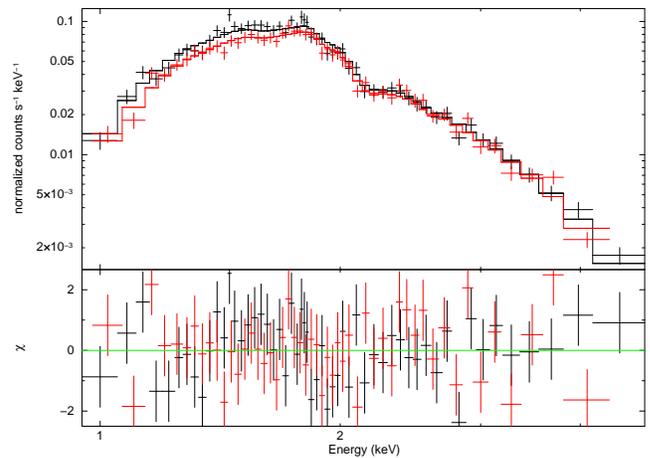}} 
\caption{The data and our best hydrogen atmosphere (NSA) model fit in the case of varying emission areas and different $N_{\rm H}$ values for the first epoch (black) and the second epoch (red). The lower panel shows the fit residuals in units of sigmas. See Table~\ref{table:NSAfits} for more details on the fit.
\label{nsadatres}}
\end{figure}
\begin{figure}[t]
{\includegraphics[width=85mm, bb=25 12 570 545, clip]{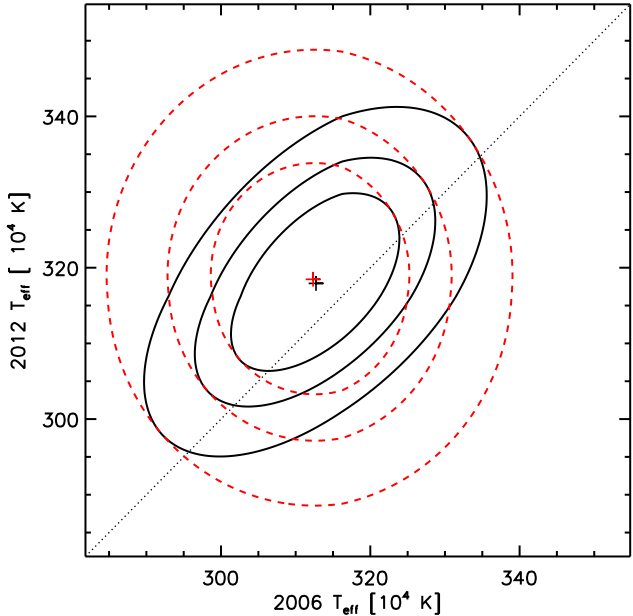}} 
\caption{Temperature confidence contours (68\,\%, 90\,\%, 99\,\%) for our NSA model fits with varying emission areas (see Table~\ref{table:NSAfits}). The black solid contours correspond to the fit where the $N_{\rm H}$ is set to be the same for both epochs, the red dashed contours correspond to the fit where the $N_{\rm H}$ is allowed to be different in the two epochs.
\label{nsaT1T2}}
\end{figure}

\begin{deluxetable*}{lcccccccc}[h!]
\tablecaption{XSPEC NSA fit results \label{table:NSAfits}}
\tablewidth{0pt}
\tablehead{
\colhead{Data} & \colhead{$N_{\rm H}$} & \colhead{$R^2_{\rm Em}/R^{2}_{\rm NS}$} & \colhead{$T_{\rm eff}$} & \colhead{$\triangle T_{\rm eff}$} & \colhead{$F^{\rm{abs}}_{-13}$} & \colhead{$F^{\rm{unabs}}_{-12}$}  & \colhead{$L_{\rm bol}^{\infty}$} & \colhead{$\chi^2_{\nu}$/d.o.f.} \\
\colhead{ } & \colhead{$[10^{22}$\,cm$^{-2}]$} & \colhead{} & \colhead{[$10^4$\,K]} & \colhead{[$10^4$\,K]} & \colhead{} & \colhead{} & \colhead{$[10^{33}$\,erg\,s$^{-1}]$} &\colhead{}
}
\startdata
\\
\multicolumn{9}{c}{{{Normalizations tied}}}\\
\\
\tableline
\\
 O1  & $2.01^{+0.10}_{-0.09}$ & $0.086^{+0.021}_{-0.014}$ & $317\pm {11}$ & &  $7.35^{+0.20}_{-0.19}$ & $2.3\pm{0.1}$ & $3.6^{+0.1}_{-0.2}$ & 1.08/102  \\[0.7ex]

O2  & $=N_H$(O1) & $=R^2_{\rm Em}/R^{2}_{\rm NS}$(O1) & $314^{+10}_{-11}$ & $-3 \pm 2$ &  $7.02 \pm {0.17}$ & $2.2 \pm 0.1$  &  $3.5 \pm 0.1$ & 1.08/102   \\[0.7ex]
  & & &  & $(-1.0$\,\% $T_{\rm O1}$) &  ($-4.5$\% $F^{\rm abs}_{\rm O1}$) & ($-5.7$\% $F^{\rm unabs}_{\rm O1}$) \\
 
\\
O1  &  $1.99^{+0.11}_{-0.09}$ & $0.087^{+0.021}_{-0.021}$ & $316 \pm 11 $ &  &  $7.35^{+0.20}_{-0.19}$  &
$2.3 \pm 0.2$ &  $3.6^{+0.3}_{-0.1}$ & 1.08/101\\[0.7ex]

O2  &  $2.04^{+0.12}_{-0.10}$ & $=R^2_{\rm Em}/R^{2}_{\rm NS}$(O1) & $314^{+10}_{-11} $ &  $ -2 \pm 3$ &  $7.02\pm 0.18$ & $2.2 \pm 0.1$ & $3.5^{+0.3}_{-0.1}$  & 1.08/101 \\[0.7ex]
  & & &  &  $(-0.5$\,\% $T_{\rm O1}$) &  ($-4.5$\% $F^{\rm abs}_{\rm O1}$) & ($-4.7$\% $F^{\rm unabs}_{\rm O1}$) \\

\\
\tableline
\\
\multicolumn{9}{c}{{{Normalizations free and untied}}}\\
\\
\tableline
\\
O1  & $2.01^{+0.10}_{-0.09}$ & $0.093^{+0.025}_{-0.017}$ & $313\pm {12}$ & &  $7.36^{+0.17}_{-0.16}$ & $2.3^{+0.2}_{-0.1}$ & $3.7^{+0.3}_{-0.2}$ & 1.08/101  \\[0.7ex]
O2  & $=N_H$(O1) & $0.081^{+0.021}_{-0.015}$ & $318 \pm 13$ & $+5 \pm 12$ &  $7.04^{+0.16}_{-0.17}$ & $2.2 \pm 0.1$  &  $3.5^{+0.3}_{-0.1}$ & 1.08/101   \\[0.7ex]
  & & &  & $(+1.7$\,\% $T_{\rm O1}$) &  ($-4.3$\% $F^{\rm abs}_{\rm O1}$) & ($-4.5$\% $F^{\rm unabs}_{\rm O1}$) \\
 
\\
O1  &  $2.01^{+0.14}_{-0.11}$ & $0.093^{+0.033}_{-0.021}$ & $312^{+14}_{-15}$ &  &  $7.36^{+0.16}_{-0.17}$  &
$2.3 \pm 0.2$ &  $3.7^{+0.5}_{-0.3}$ & 1.09/100\\[0.7ex]
O2  &  $2.00^{+0.16}_{-0.14}$ & $0.080^{+0.031}_{-0.020}$ & $318 \pm 17 $ &  $+6 \pm 22$ &  $7.04\pm 0.17$ & $2.2 \pm 0.2$ & $3.4^{+0.5}_{-0.3}$  & 1.09/100 \\[0.7ex]
  & & &  &  $(+2.0$\,\% $T_{\rm O1}$) &  ($-4.3$\% $F^{\rm abs}_{\rm O1}$) & ($-4.5$\% $F^{\rm unabs}_{\rm O1}$) \\
\enddata

\tablecomments{The fits are done simultanously for O1 and O2 (first and second observation epoch, respectively) with the same $R_{\rm NS}=10$\,km ($R^{\infty}_{\rm NS}=13.06$\,km), $M_{\rm NS}=1.4$\,M$_{\odot}$. A distance $d=3.4$\,kpc is assumed to calculate $R^2_{\rm Em}/R^{2}_{\rm NS}$ from the model normalization. $T_{\rm eff}$ is the unredshifted effective temperature. The flux values are given for an energy range of 0.6-6\,keV, the absorbed fluxes are in units of $10^{-13}$\,erg\,cm$^{-2}$\,s$^{-1}$, while the unabsorbed fluxes are in units of $10^{-12}$\,erg\,cm$^{-2}$\,s$^{-1}$.  
All errors indicate the 90\% confidence level for one parameter of interest.
The bolometric luminosity at inifinity is calculated as $L^{\infty}_{\rm bol}=4 \pi \sigma {R^{\infty}_{\rm Em}}^2 {T^{\infty}_{\rm eff}}^4$. 
The luminosity  and temperature uncertainties were determined from the 90\% confidence contours for one parameter of interest, which are a factor 0.77 the contours for two parameters of interest in Figures~\ref{fig:normNHT} and \ref{nsaT1T2}.
}
\end{deluxetable*}

We used the NS hydrogen atmosphere models \citep{Pavlov1995,Zavlin1996}, \texttt{NSA} in XSPEC; to fit the data of the two epochs simultaneously. Similarly to PL09, the NSA models with high magnetic fields ($B=10^{12}$\,G or $B=10^{13}$\,G) result in worse fits (reduced $\chi^2_{\nu}=1.4$ and 1.5 for $\nu=101$ degrees of freedom) than the NSA model with a low magnetic field ($B<10^{10}$\,G, $\chi^2_{\nu}=1.1$). The results of the latter are given in Table~\ref{table:NSAfits}. 

\begin{figure}[t]
{\includegraphics[width=85mm, bb=25 12 570 540, clip]{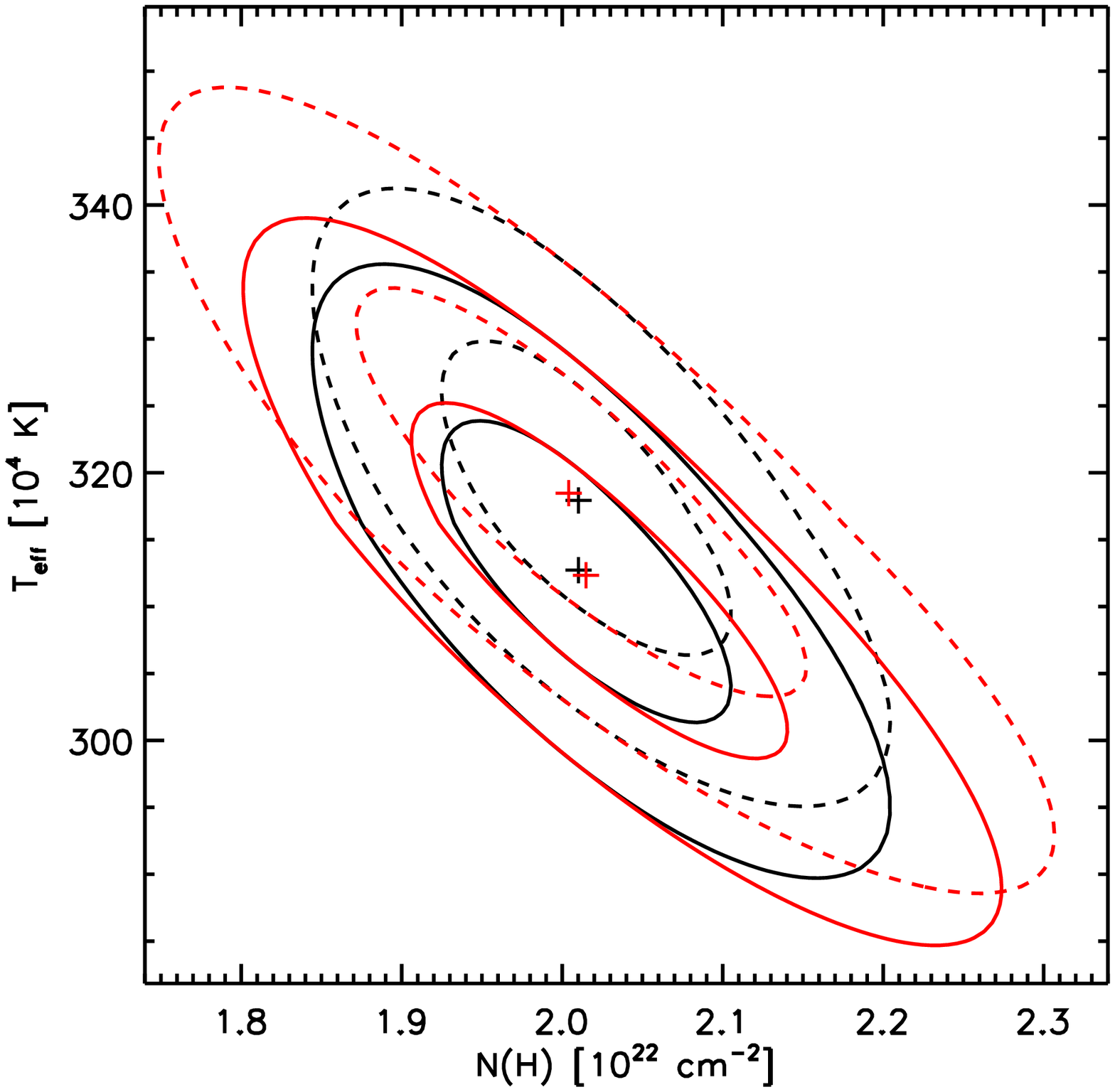}} 
\caption{Temperature versus $N_{\rm H}$ confidence contours (68\,\%, and 99\,\%) for our NSA model fits allowing varying emission areas (see Table~\ref{table:NSAfits}). The black contours mark the fit where $N_{\rm H}$ is set to be the same for both epochs, the red contours mark the fit where $N_{\rm H}$ is allowed to be different in the two epochs. Solid lines indicate contours of the 2006 data, dashed lines contours of the 2012 data. 
\label{fig:nsaNHT}}
\end{figure}

\begin{figure}[t]
{\includegraphics[width=85mm, bb=25 12 570 540, clip]{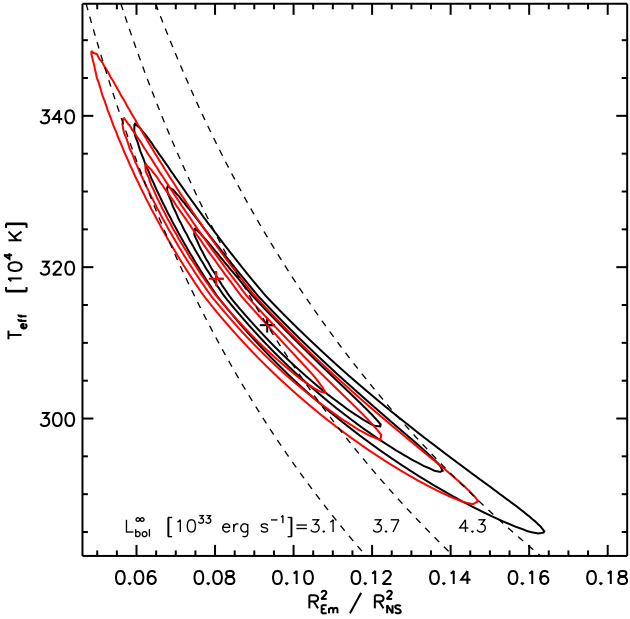}} 
\caption{Confidence contours (68\,\%, 90\,\%, 99\,\%) of our NSA model fits (see Table~\ref{table:NSAfits}) for the temperature versus the ratio of the effective emission area to the surface area of a $R_{\rm NS}=10$\,km neutron star. $N_{\rm H}$ and the emission areas are allowed to be different in the two epochs. Black lines indicate contours of the 2006 data, red lines contours of the 2012 data. The dashed lines are the lines of constant bolometric luminosity at inifinity, $L^{\infty}_{\rm bol}=4 \pi \sigma {R^{\infty}_{\rm Em}}^2 {T^{\infty}_{\rm eff}}^4$. 
\label{fig:normNHT}}
\end{figure}

\begin{deluxetable*}{lccccccc}[t]
\tablecaption{Fit results for a modified thickness of the ACIS contaminant, using carbon atmosphere models with $\log g=14.45$ and $z=0.375$ with fixed normalizations in both epochs \label{table:carbonfitscontam}}
\tablewidth{0pt}
\tablehead{
\colhead{Data} & \colhead{$\triangle l/l$} & \colhead{$N_{\rm H}$} & \colhead{$T_{\rm eff}$} & \colhead{$\triangle T_{\rm eff}$} & \colhead{$F^{\rm{abs}}_{-13}$} & \colhead{$F^{\rm{unabs}}_{-12}$}  & \colhead{$\chi^2_{\nu}$/d.o.f.} \\
\colhead{ }  & \colhead{} & \colhead{$[10^{22}$\,cm$^{-2}]$} & \colhead{[$10^4$ K]} & \colhead{[$10^4$ K]} & \colhead{} & \colhead{} & \colhead{}   
}
\startdata
\multicolumn{8}{c}{{{the same $N_{\rm H}$ in both epochs }}}\\
\tableline
\\
O1&   &  $2.26 \pm 0.05$ & $201.2^{+1.1}_{-1.0}$ &  $\cdots$ & $7.35 \pm 0.17$  & $2.92 \pm 0.10$ & 1.058/103\\[0.7ex]
O2 &  $+10$\% & $=N_H$(O1) & $199.9^{+1.0}_{-1.1}$ & $-1.4 \pm 1.1$ & $7.03 \pm 0.17$  & $2.81^{+0.10}_{-0.09}$ & 1.058/103    \\[0.7ex]
& &  &  & ($-0.7$\% $T_{\rm O1}$) &    \\
\\
O1&   &  $2.23 \pm 0.05$ & $200.9^{+1.0}_{-1.1}$ & $\cdots$ & $7.32 \pm 0.17$  & $2.88 \pm 0.10$ & 1.065/103\\[0.7ex]
O2 &  $+30$\%  & $=N_H$(O1) & $200.2^{+1.0}_{-1.1} $ & $-0.7\pm 1.1$ & $7.18 \pm 0.17$  & $2.84^{+0.10}_{-0.09}$ & 1.065/103    \\[0.7ex]
& &  &  & ($-0.3$\% $T_{\rm O1}$) &    \\
\\
O1 &    &  $2.29 \pm 0.05$ & $201.6 \pm 1.0$ & $\cdots$ & $7.38 \pm 0.17$  & $2.95 \pm 0.10$ & 1.071/103\\[0.7ex]
O2 &  $-10$\%  & $=N_H$(O1) & $199.5 \pm 1.1$ & $-2.1\pm 1.1$ & $6.89 \pm 0.17 $  & $2.78^{+0.10}_{-0.09}$ & 1.071/103    \\[0.7ex]
& &  &  & ($-1.1$\% $T_{\rm O1}$) &    \\
\\
O1 &    &  $2.31 \pm 0.05$ & $202.0^{+1.1}_{-1.0}$ & $\cdots$ & $7.41 \pm 0.17$  & $2.98 \pm 0.10$ & 1.107/103\\[0.7ex]
O2 &  $-30$\%  & $=N_H$(O1) & $199.1 \pm 1.1$ & $-2.9 \pm 1.1$ & $6.75 \pm 0.17$  & $2.75^{+0.10}_{-0.09}$ & 1.107/103    \\[0.7ex]
& &  &  & ($-1.4$\% $T_{\rm O1}$) &    \\
\\
\tableline
\multicolumn{8}{c}{{{ different $N_{\rm H}$ in the two epochs }}}\\
\tableline
\\
O1  &  & $2.25 \pm 0.06$ & $201.1 \pm 1.2$ & $\cdots$ & $7.34 \pm 0.18$  & $2.90^{+0.12}_{-0.11}$ & 1.066/102\\[0.7ex]
O2 & $+10$\%  &  $2.27 \pm 0.07$ & $200.1^{+1.3}_{-1.4}$ & $-1.0\pm 1.8$ & $7.04 \pm 0.18$ & $2.82\pm 0.12$ & 1.066/102 \\[0.7ex]
 & &  &  & ($-0.5$\% $T_{\rm O1}$) & \\
\\ 
O1 &   &  $2.25 \pm 0.06$ & $201.1 \pm 1.2 $ & $\cdots$ & $7.34 \pm 0.18$  & $2.90^{+0.12}_{-0.11}$ & 1.072/102\\[0.7ex]
O2&  $+30$\%  &  $2.21 \pm 0.07$ & $199.9^{+1.3}_{-1.4} $ & $-1.1\pm 1.8$ & $7.15 \pm 0.18$ & $2.81 \pm 0.12$ & 1.072/102 \\[0.7ex]
 &  &   &  & ($-0.6$\% $T_{\rm O1}$) & \\
\\
O1 &   &  $2.25 \pm 0.06$ & $201.1 \pm 1.2 $ & $\cdots$ & $7.34 \pm 0.18$  & $2.90 \pm 0.12$ & 1.060/102\\[0.7ex]
O2& $-10$\%   &  $2.33 \pm 0.07$ & $200.2^{+1.3}_{-1.4} $ & $-0.9 \pm 1.8$ & $6.94 \pm 0.18$ & $2.83 \pm 0.12$ & 1.060/102 \\[0.7ex]
 &  &   &  & ($-0.4$\% $T_{\rm O1}$) & \\
\\ 
O1  &   &  $2.25 \pm 0.06$ & $201.1 \pm 1.2 $ & $\cdots$ & $7.34 \pm 0.18$  & $2.90^{+0.12}_{-0.11}$ & 1.055/102\\[0.7ex]
O2 &  $-30$\% &  $2.39 \pm 0.07$ & $200.3^{+1.3}_{-1.4} $ & $-0.7 \pm 1.8$ & $6.83 \pm 0.17$ & $2.84 \pm 0.12$ & 1.055/102 \\[0.7ex]
 &   &  &  & ($-0.4$\% $T_{\rm O1}$) & \\
\enddata

\tablecomments{The same definitions as in Table~\ref{table:carbonfits} are used in this table.
The effective area of the second epoch has been modified to check different contamination uncertainty ranges as described in Appendix~\ref{contamtest1}. The change of the thickness of the contamination layer is indicated in the column $\triangle l/l$.
For example, ``O2$+10$\,\%'' indicates an increase of the contamination layer thickness, thus, its optical depth by 10\%, the same for all photon energies. 
All fits were done for fixed normalizations, $\mathcal{N}=923$, which corresponds to the distance of 3.4\,kpc assuming emission from the whole uniformly heated surface of a neutron star with $\log g=14.45$ and $z=0.375$.
As in Table~\ref{table:carbonfits}, all errors indicate the 90\% confidence level for one parameter of interest.
The temperature differences were calculated using the temperature values before rounding. The errors of the temperature difference were determined from the 90\% confidence contours.} 
\end{deluxetable*}

In addition to the small magnetic field, we used the model with fixed $R_{\rm NS}=10$\,km ($R^{\infty}_{\rm NS}=13.06$\,km) and  $M_{\rm NS}=1.4$\,M$_{\odot}$ ($\log g=14.39$, $z=0.306$). As outlined in Section~\ref{resultscarbon}, we cannot exclude $N_{\rm H}$ variability.
Therefore, we checked whether allowing $N_{\rm H}$ to vary between the observations gives different fit values but found that the change of $N_{\rm H}$ is much smaller than the $N_{\rm H}$ uncertainties.
The fit in the case of different $N_{\rm H}$ and varying emission area sizes is shown in Figure~\ref{nsadatres}.
In general, low magnetic field NSA models fit the spectra of both epochs well. Even better fits are obtained for the two-component NSA+NSA models (see PL09). Here, we do not investigate these two-component models because the temperature uncertainties would be too large to address any temperature change.  We assume a distance of $d=3.4$\,kpc to calculate the ratio $R^2_{\rm Em}/R^{2}_{\rm NS}$ from the NSA normalization, where $R^2_{\rm Em}$ is the effective emission radius. 
The temperatures, emission areas and unabsorbed fluxes are the same for both epochs, considering fit uncertainties (see Table~\ref{table:NSAfits}). 
The best-fit absorbed and unabsorbed flux decreases in the energy range 0.6-6\,keV are $\sim 4$\% to  $\sim 6$\% but their significance does not exceed the $3\sigma$ level. If the same emission area sizes are assumed in both epochs, the flux decrease is formally due to a slight temperature decrease. If instead the emission area is allowed to change, the 
slight decreases of the best-fit fluxes can be attributed to the decrease of the best-fit emission areas.
The seen {\emph{in}}creases of the best-fit temperatures in this case are within the 68\,\% confidence levels (Figure~\ref{nsaT1T2} and Table~\ref{table:NSAfits}, one parameter of interest for the considered temperature difference).
The 68\,\% confidence contours (two parameters of interest) of both epochs also overlap in the temperature versus $N_{\rm H}$ plot (Figure~\ref{fig:nsaNHT}) and the temperature versus emission area plot (Figure~\ref{fig:normNHT}).     
All the inferred emission areas are, however, a factor of $\sim 10$ smaller than the NS surface area.

\subsection{The effect of the ACIS contamination uncertainty}
To investigate the impact of the uncertainty in the optical depth of the ACIS contamination, we used the carbon atmosphere models. Applying modified effective areas for the second epoch produced as described in the Appendix~\ref{contamtest1}, we used the $\log g=14.45$, $z=0.375$ carbon atmosphere model with fixed normalizations in the fits of the two epochs. The results are listed in Table~\ref{table:carbonfitscontam}.\\
\begin{figure}[b]
{\includegraphics[width=85mm, bb=25 12 590 550, clip]{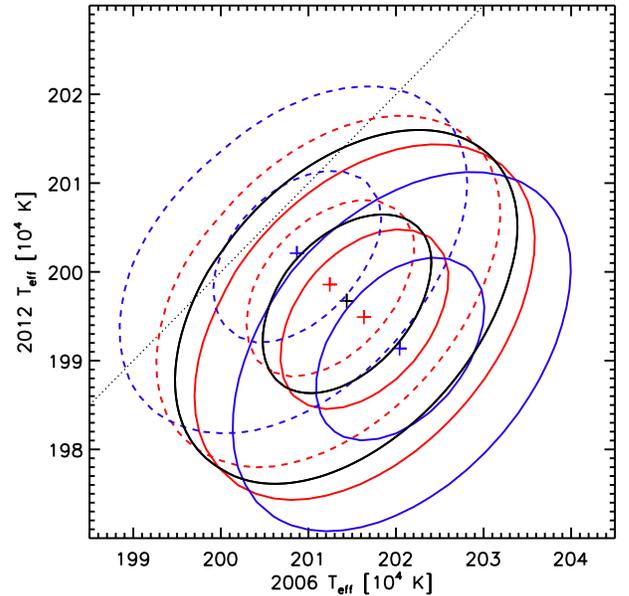}} 
\caption{Temperature confidence contours (68\,\% and 99\,\%) for the fit to the ($\log g=14.45$, $z=0.375$) carbon atmosphere model using different thicknesses of the contamination layer on the ACIS detector, which correspond to the same relative changes in the optical depth. The black contours show the original fit obtained with the standard contamination correction in CIAO (same contours as in Figure~\ref{carbonT1T2zoom}). 
The red contours indicate changes of the contamination optical depth by 10\,\%, blue contours indicate changes of 30\,\%. Dashed lines mark increasing optical depths, solid lines mark decreasing optical depths. 
The $N_{\rm H}$ is set to be the same for both epochs in all fits.
\label{carbonT1T2contam}}
\end{figure}

In the case of the same $N_{\rm H}$ in both epochs, a larger optical depth (i.e., increased thickness in 2012) of the contamination layer leads to smaller $N_{\rm H}$, temperature and flux differences (see Table~\ref{table:carbonfitscontam} and Figure~\ref{carbonT1T2contam}). A decreased contamination layer thickness in 2012 results in the opposite effect.
Compared to the original fit, the $\chi^2_{\nu}$ marginally improves for $\triangle l/l=+10$\%, where $\triangle l/l$ is the change of the thickness of the contamination layer. The $\chi^2_{\nu}$ is worse for all other (larger or smaller) contamination optical depths.  
For a reasonable uncertainty of the contamination layer, $\pm 10$\,\%, the found  range of temperature change is $\triangle T_4=\triangle T_{\rm eff} / 10^{4}$\,K$=-2.1 \pm 1.1$ to $\triangle T_4=-1.4 \pm 1.1$ (90\% confidence errors).
In the case of the 10\% decreased contamination layer, the temperature change is  at the $3.1\sigma$ level.\\
 
Thus, if the contamination layer in 2012 is {{over}}estimated by at least 10\% in the current calibration models, the temperature decrease would be significant in the case of the same $N_{\rm H}$ in the two epochs.
For an uncertainty of $\pm 10$\,\% of the 2012 contamination layer, the \emph{unabsorbed} fluxes are consistent with each other within their 90\% confidence errors. In the case of a 10\% thicker or thinner contamination layer, the significance of the \emph{absorbed} flux drop reaches the $2.3\sigma$ level and $3.5\sigma$ level, respectively.\\

In the case of a variable $N_{\rm H}$ in the two epochs, the effect of a changing contamination layer on the temperature difference is smaller since the second $N_{\rm H}$ can partly account for the changed effective area.  
In fact, the effect on the temperature is in the opposite direction compared to the case of the same $N_{\rm H}$  -- the temperature difference gets larger or smaller for an  increasing  or decreasing contamination optical depth, respectively. 
Compared to the original fit of the variable-$N_{\rm H}$ case, the $\chi^2_{\nu}$ marginally improves for decreasing contamination.
For a reasonable uncertainty of the contamination layer, $\pm 10$\,\%, the temperature and flux differences are less significant than in the case of the same $N_{\rm H}$, all difference values are below the $3\sigma$ level (see Table~\ref{table:carbonfitscontam}).

\subsection{Timing Analysis}
We did a similar timing analysis for the 2012 data as done by PL09 for the 2006 data. 
We extracted events in a circle around the CCO centroid position with a radius of 2.9\,pixels, obtaining 6311 total counts which includes 508 background counts for the 2012 observation in the energy range of 0.3 to 8\,keV. The times-of-arrival of the X-ray photons were corrected to the solar system barycenter system using the CIAO tool \texttt{axbary}.
We calculated the $Z_1^2$ statistic (e.g., \citealt{Buccheri1983}) to search for pulsations of the CCO.
Considering the frame time, $t_{\rm frame}=0.34104$\,s, we searched in the period range of 0.68\,s to 20\,s (or a frequency range of  $0.05$\,Hz$<\nu <1.47$\,Hz). Using a sampling of 0.5\,$\mu$\,Hz, we did not find any significant $Z_1^2$ peak in the data ($Z^2_{1,\rm{max}} = 26.4$  at $\nu=0.958305$\,Hz has a chance probability $p= 0.19$). 
Similarly to PL09, we follow the approach by \citet{Groth1975} to estimate the upper limits on the intrinsic pulsed fraction for the period range of 0.68\,s to 20\,s as 13\%, 15\% and 16\%  at the confidence levels of 95\%, 99\%, 99.9\%. 

\section{Discussion}
In this section we first discuss the consistency of our analysis methods and results on the 2006 data to previous ones (Section~\ref{disprev}). This is followed by a discussion of our results for the temperature and flux differences (Section~\ref{distemp}). In the end we briefly discuss the applicability of different atmosphere models to the Cas A CCO (Section~\ref{disatmos}).

\subsection{Comparison with previous works on the 2006 data}
\label{disprev}
Detailed analyses of the 2006 data were presented by PL09 and HH09. Temperature fit results were also reported by HH10 for the 2006-only fit where the other parameters were fixed at values derived from the fits of the ACIS-S3 piled-up Graded mode data. We concentrate in the following on these three works.\\ 

Compared to the previously published results, all our 2006-only fits show a significantly higher hydrogen column density. In the case of the NSA model fits by us and PL09, the discrepancy of $\triangle N_{\rm H,22}=0.44$, where $N_{\rm H,22}$ is in units of  $10^{22}$\,cm$^{-2}$, can be explained by the different abundances and effective cross sections used in calculating the photoelectric absorption. Applying \texttt{wabs} and the abundance tables by \citet{Anders1989} instead of the newer ones by \citet{Wilms2000}, the obtained $N_{\rm H,22}=1.48^{+0.08}_{-0.07}$ agrees within errors with the $N_{\rm H,22}=1.57^{+0.08}_{-0.10}$ by PL09.\\ 

Our $N_{\rm H}$ values ($N_{\rm H,22} \approx 2.3$) are, however, also significantly larger than the values by HH09 for both the hydrogen and carbon atmosphere models ($N_{\rm H,22} \approx 1.7$), although HH09 used the same absorption model \texttt{tbabs} and abundances.
Apart from the fact that most of the observations used by HH09 suffer from pile-up, and are telemetered in Graded mode, two other main differences are the selection of the background region and the accounting for dust scattering with a fixed gas-to-dust ratio.
Extracting the same source and background regions as HH09, and applying the XSPEC `dust' scattering model, we can reproduce $N_{\rm H}$ values consistent with the values obtained by HH09. 
As explained in the Appendix~\ref{appdust}, we neglect dust scattering in the rest of the paper because the applied simplifyed model has no significant influence on the studied temperature difference.\\

For the hydrogen atmosphere model fits, the other parameters in our NSA fit results are consistent within errors with the corresponding values reported by PL09 (using the same abundances for both fits).
Comparing our hydrogen atmosphere best-fit parameters for 2006 to those of HH09, we find a slight temperature difference and a difference in the fraction of the X-ray emitting NS surface area ($0.09^{+0.03}_{-0.02}$ from Table~\ref{table:NSAfits} versus $0.18 \pm 0.03$ from HH09) at goodness of fit indicators comparable to our values. We checked that the seen small differences can be mostly attributed to different choices of background regions. The additional use of the megasecond long Graded mode observation of the Cas\~A SNR by \citet{Hwang2004} had also an impact on the HH09 fit results.\\

For the carbon atmosphere model fits, HH09 reported a NS surface temperature of $T_4=180^{+8}_{-10}$ (joint fit of the 2006 and the \citealt{Hwang2004} data), while HH10 reported for the 2006 data a surface temperature of $T_4=203.2^{+0.4}_{-0.7}$ ($1\sigma$ errors), where the values were derived for a NS with  $\log g=14.4534$, $z=0.377$.
Our fit to a carbon atmosphere model with very similar gravitational parameters ($\log g=14.45$, $z=0.375$, dust scattering included) resulted in a comparable surface temperature, $T_4=201.3 \pm 0.7 (1\sigma) $. \\

We conclude that the fit results using our carbon atmosphere models are consistent with those by HH10 for data {\emph{with negligible pile-up}} taken in Faint mode. Thus, the two independent sets of carbon atmosphere models produce comparable results in the parameter range of interest.\\

\subsection{Temperature and flux differences}
\label{distemp}
As shown in Section~\ref{resultsnsa} and listed in Table~\ref{table:NSAfits}, there is no significant temperature decrease if we fit the data with the hydrogen atmosphere models. Depending on whether the emission areas are tied or not in the two epochs, we obtain a slightly higher or lower best-fit temperature for the second epoch, respectively; however, the difference in each case is statistically insignificant (e.g., $\triangle T_4=+6\pm 22 $, for different $N_{\rm H}$ and emission areas). 
An apparent $\sim 4$\% flux decrease can be formally attributed to the decrease of the emitting area or of the temperature for varying or tied emission areas, respectively. 
However, the significance of the model flux decrease is below the $3\sigma$ level. 

The temperature and the (small) emission area are strongly correlated, but the fits to the 2006 data and the 2012 data overlap within their 68\% confidence contours (Figure~\ref{fig:normNHT}). Thus, the bolometric luminosities of the hydrogen atmosphere models are consistent within their errors as well (see Table~\ref{table:NSAfits}).\\

In the case of the carbon atmosphere fits (Section~\ref{resultscarbon} and 
Table~\ref{table:carbonfits}), the obtained  best-fit temperature in 2012 is lower than the one in 2006 if the normalizations are fixed or tied, but it is higher if both normalizations are free fit parameters. Since the statistical significance of the temperature drop over the time span of 5.5\,years does not exceed $3\sigma$ ($1\sigma$ in the case of varying $N_{\rm H}$, which we consider more realistic), we can only estimate an upper limit on the drop.
As a conservative estimate, we define the upper limit as the sum of the best-fit drop and its 90\% uncertainty. Such upper limits are in the range of $-\triangle T_4< 3.1$ (the same $N_{\rm H}$ in both epochs, tied $\mathcal{N}$) and  $-\triangle T_4< 2.7$ (different $N_{\rm H}$, the same fixed $\mathcal{N}$ in both epochs), for the default contamination. The upper limit slighty increases if we include the additional uncertainty of the contamination thickness ($-\triangle T_4< 3.2$ for contamiantion changes at the 10\% level).\\ 

The question arises whether we can completely exclude -- for our time baseline and data -- a temperature drop on the order of what has been found by HH10 and E+13.   
HH10 reported an overall surface temperature decrease of 3.6\,\%$\pm 0.6$\,\% over a time span of 9.8 years. 
Using an enlarged data set and a CTI correction in the Graded mode, 
E+13 found  3.5\,\%$\pm 0.4$\,\% (from 2000 to 2010) and estimated an additional systematic uncertainty due to the choice of their background as $(+1.6$\%, $-0.3$\%).\\ 

\begin{figure}[b]
{\includegraphics[width=85mm, bb=20 14 682 542, clip]{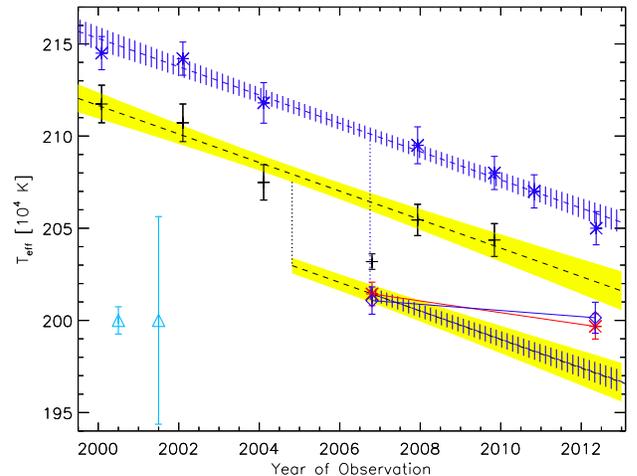}}
\caption{ Temperature change over time. The black crosses and blue asterisks mark the temperatures and their $1\sigma$ errors as reported by HH10 (their Table 1 and Figure 2) and E+13 (their Table 2), respectively. 
Their fit results were derived from piled data telemetered in Graded mode, using carbon atmosphere models ($M_{\rm NS}=1.648$\,M$_{\odot}$ and $R_{\rm NS}=10.3$\,km (HH10); $M_{\rm NS}=1.62$\,M$_{\odot}$ and $R_{\rm NS}=10.19$\,km; E+13) with the same fixed $N_{\rm H}$ for all observations.
The black dashed line and the yellow area indicate the results of a linear regression fit and its $1\sigma$ error to the HH10 data points if we choose the average of their observing epochs as reference time, $t_{\rm HH10,0}=2004.82$ (dotted vertical black line).   
The blue dashed line and the blue-striped area indicate the results of a linear regression fit and its $1\sigma$ error to the E+13 data points if we choose the average of their observing epochs as reference time, $t_{\rm E13,0}=2006.75$ (dotted vertical blue line).   
Our fit results from Table~\ref{table:carbonfits} for a carbon atmosphere model with  similar gravitational parameters ($M_{\rm NS}=1.647$\,M$_{\odot}$ and  $R_{\rm NS}=10.33$\,km) are marked with red star points (same $N_{\rm H}$ in both epochs) and blue, diamond points (different $N_{\rm H}$ in 2006 and 2012).  For completeness, we also show the 2006 temperature by HH10.  All errors in this plot are $1\sigma$ errors. In the lower left corner, we show a typical $1\sigma$ temperature uncertainty  for the cases of fixed (left) and free, but tied (right) normalizations. See text for a detailed discussion.
\label{tempfit}}
\end{figure}

To obtain an average yearly temperature change rate from the piled data points and the respective errors presented by HH10, we performed a standard least-square fit to a straight line (e.g., \citealt{Bevington2003}), $T_{\rm eff}= T_0 + \dot{T}(t-t_0)$, where $t$ is the time and  $t_0$ the reference time. 
We chose the average value of their covered time span as the zero-point of the independent variable, $t_0=2004.816$. 
We derived a slope
$\dot{T}=-7700\pm 1900$\,K\,yr$^{-1}$ and an intercept $T_0=(207.9 \pm 0.7)\times 10^4$\,K (all errors indicate 90\% confidence levels, $\chi^2_{\nu}=0.56$ for $\nu=3$\,d.o.f.), shown in yellow in Figure~\ref{tempfit}. 
We followed the same approach for the data points presented by E+13 and obtained $\dot{T}=-7700\pm 1300 $\,K\,yr$^{-1}$ and $T_0=(210.1 \pm 0.6)\times 10^4$\,K ($\chi^2_{\nu}=0.41$ for $\nu=5$\,d.o.f.), indicated by blue stripes in Figure~\ref{tempfit}. 
For illustration, assuming constant (time-independent) systematic shifts between the fit results of the Graded mode data by HH10 and E+13 with respect to our results, we shifted the straight-line fits by constant values in such a way that the fit predictions at the time of our first epoch are going through the value of our fit result.\\

Obviously, there are several systematic errors involved in this simple comparison. 
The employed systematic shifts for our linear-regression fits could be larger or smaller.
In fact, the systematic shift can be different for each observation due to the different locations of the target on the detector with account for the spatial dependencies of the pile-up and ACIS contamination.  Taking these uncertainties into account, one can expect larger error bars for any `shifted' data points, hence the total uncertainty of any fit to these data will be larger. 
The different contributions to the systematic errors from the poorly modeled pile-up and the CTI correction in Graded mode are unknown and cannot be assessed without directly comparable observations.\\

The uncertainties in the temperature difference are relatively large, in particular if we take into account the possible spread in values due to the uncertainty in the ACIS filter  contamination (Tables~\ref{table:carbonfits} and \ref{table:carbonfitscontam}, Figures~\ref{carbonT1T2big}, \ref{carbonT1T2zoom}, and \ref{carbonT1T2contam}).
Calculating the average temperature change {{per year}} for our data, we obtain a range of $\dot{T}$ from $-1600\pm 3200$\,K\,yr$^{-1}$ to $-1800\pm 3200$\,K\,yr$^{-1}$ (90\% confidence levels) for varying $N_{\rm H}$, fixed $\mathcal{N}$, and considering a $\pm 10$\,\% uncertainty in the optical depth of the ACIS contaminant (Tables~\ref{table:carbonfits} and \ref{table:carbonfitscontam}).
If one sets $N_{\rm H}$ to be the same in the observing epochs, $\dot{T}$ ranges from $-2500\pm 2000$\,K\,yr$^{-1}$ to  $-3800 \pm 2000$\,K\,yr$^{-1}$ (90\% confidence levels, fixed $\mathcal{N}$ ), respectively, i.e., the yearly change is still considerably smaller, and its error is larger than those found from the HH10 and E+13 results.
Only for the same $N_{\rm H}$ and 10\% $less$ contamination in 2012, the values barely overlap at the 90\% confidence levels.
If we consider, however, the ranges at the 99\% confidence levels, all slopes overlap.  Thus, we cannot firmly exclude a temperature change on the order of what has been reported before by E+13 (or HH10), but the probability that this temperature change is correct appears to be rather low.\\

The most likely reason of the discrepancy in $\dot{T}$ is the lower quality of the data used by HH10 and E13, subject to strong pile-up and other calibration issues (e.g., the effects of the CTI cannot be reliably corrected for the Graded-mode data). We suspect that these  calibration issues had an impact on the data analyzed by HH10 and E+13, leading to an underestimate of the error in the previously reported temperature decline\footnote{Note that the additional systematic error reported by E+13 takes only the background choice into account, but not the uncertainties of pile-up and CTI corrections.}.
As shown in our analysis, it furthermore matters whether one assumes that $N_{\rm H}$ is the same or variable in the different epochs (HH10 and E+13 used the same \emph{fixed} $N_{\rm H}$ in all their spectral fits). 
The fit uncertainties of our $N_{\rm H}$ values in the two epochs are large and of the same order as the obtained $N_{\rm H}$ difference (e.g., Tables~\ref{table:carbonfits} and \ref{table:carbonfitscontam}). Therefore, it is currently not possible to say whether  $N_{\rm H}$ has indeed changed or not. However, given the dynamic and inhomogeneous environment of the SNR and the ISM around it (see Section~\ref{resultscarbon}),
a varying $N_{\rm H}$ appears to be a realistic choice, in particular if the same background regions are used in all epochs.\\   

Our upper limit on the cooling rate of the CCO between 2006 and 2012 is $-\dot{T}< 5100$\,K\,yr$^{-1}$ (calculated as the sum of the best-fit yearly rate plus its 90\% confidence uncertainty) for fixed $\mathcal{N}$, varying $N_{\rm H}$, and with account for the uncertainty in the optical depth of the ACIS contamination layer.
In the case of the tied (but fitted) $\mathcal{N}$ and varying $N_{\rm H}$, the upper limit on the cooling rate of the CCO between 2006 and 2012 is $-\dot{T}< 5400$\,K\,yr$^{-1}$ (default contamination).
With the likely future improvements of the contamination model, our upper limit could be further refined in the future.\\

The {{unabsorbed}} fluxes of 2006 and 2012 are consistent with each other within their 90\% confidence levels for all the models (varied and tied $N_{\rm H}$ and $\mathcal{N}$, $\pm 10$\,\% uncertainty in the ACIS contamination layer thickness). 
The change of the absorbed flux, however, is more significant, approaching the $3\sigma$ level.
In Figures~\ref{ufflux}, we showed that there is a hint of a flux decrease in a particular energy region ($\approx 1.4-1.8$\,keV).
If this energy-restricted flux decrease is real, it could indicate problems in the ACIS filter contamination model. Currently, the model assumes the same accumulation rate for the individual contaminant components. If the contaminants accumulate differently, one would expect such an apparent energy-restricted flux change. 
Another explanations could be unresolved patches of matter in the SNR (e.g., cold dust clumps, see also Section~\ref{resultscarbon}) moving across the sight line to the CCO. 
In order to test whether the flux decrease is a calibration issue, one could reobserve soft non-varying X-ray sources with ACIS. Good calibration sources are most of the X-ray thermal isolated NSs which are also monitored with the stable XMM-$Newton$ EPIC pn instrument (e.g., \citealt{Sartore2012}), allowing for an independent check of any suspected temperature or flux changes.\\

The bolometric luminosities
are different for the NSA hydrogen atmosphere model and the carbon atmosphere model, but the respective values of the two epochs are consistent with each other at the 90\% confidence level (see Table~\ref{table:NSAfits} and Table~\ref{table:carbonfits}). 
The luminosity difference between the hydrogen and carbon atmosphere models is due to the fact that the emitting area of the former is a factor of $\sim 10$ smaller than the emitting area of the latter, but the hydrogen atmosphere temperature is only a factor $\sim 1.6$ higher than the inferred temperatures of the carbon atmosphere model fits.\\

\subsection{Is there a preferable atmosphere model ?}
\label{disatmos}
Formally, the hydrogen (NSA) and carbon atmosphere models fit the data similarly well. Although the carbon atmosphere fit has a slightly smaller $\chi^2_{\nu}$ (Tables~\ref{table:NSAfits} and \ref{table:carbonfits}), the difference is minute, and the NSA fit becomes better if an additional NSA component is included (PL09).\\

The NSA model fits imply that only a small part of the NS surface is responsible for the observed X-ray emission (for a distance of 3.4\,kpc). At a favorable viewing geometry, such hot spot emission would be expected to result in X-ray pulsations, which are currently not detected. As mentioned in Section~\ref{intro}, however, the current upper limits on the pulsed fraction of the Cas A CCO are \emph{above} the pulsed fractions seen for two of the three other CCOs with detected X-ray pulsations (and small emission areas). Therefore, the non-detection of X-ray pulsations for the Cas A CCO is not constraining with respect to the existence of small hot spots.\\ 

The origin of small hot spots is difficult to explain if only low dipole magnetic fields -- as obtained for three CCOs from timing by \citet{Gotthelf2013} and \citet{Halpern2010} -- are considered.
PL09 discussed several possible explanations.  Their most promising scenario, in the light of the recent theoretical work by \citet{Vigan2012} and \citet{Vigan2013a}, is the presence of a large toroidal magnetic field in the NS crust which would lead to a large temperature contrast between the poles and the rest of the NS surface \citep{2006perez}. The idea that a large crustal toroidal magnetic field aligned with the magnetic dipole axis can suppress heat conduction everywhere except around the magnetic poles, has initially been explored by \citet{Geppert2004, Geppert2006}.
\citet{Vigan2012} concluded from their study that the `hidden magnetar' model is a viable scenario to explain CCOs. Their two-dimensional simulations allowed them to account for the Hall drift which can generate toroidal and multipolar components in the inner crust. \citet{Vigan2012} showed that in the case of a `hidden magnetar' the reemerging field can indeed create a highly anisotropic surface temperature distribution. They proposed the difference between the isotropic and anisotropic surface temperature distribution as a possibility to distinguish between the `anti-magnetar' and the `hidden magnetic field' scenarios.
Considering these latest results together with the previous work on this subject (e.g., \citealt{Ho2011, Geppert1999}) as well as the relatively large pulsed fraction limit, we conclude that the small size of the emission region does not exclude the applicability of hydrogen atmosphere models for the Cas A CCO.\\

An argument for the use of the carbon atmosphere model is the more realistic radius inferred for the emission area (HH09). 
The mass and radius  confidence contours deduced for the carbon atmosphere model cover well the range expected for state-of-the-art NS equations of state (Appendix~\ref{carbmodels}, Figure~\ref{carbonMassRadius}), while the area inferred from the hydrogen atmosphere fits is far too small compared to the whole NS surface.
However, as we describe in the Appendix~\ref{carbmodels} and show in Figure~\ref{carbonMassRadius}, the actually inferred radius (and mass) are highly uncertain. 
Even for the carbon atmosphere model one cannot exclude the possibility that the X-ray emission comes from a fraction of the NS surface, given the uncertainty of the contours as well as the uncertainty of the distance.\\

Another indirect argument in favor of using the carbon atmosphere model for the Cas A CCO is that the spectrum of another CCO, in the SNR G353.6-0.7, can be fitted with such an atmosphere too, resulting in a reasonable radius estimate \citep{Klochkov2013}. That spectrum can, however, be also fitted with a hydrogen atmosphere model with a smaller effective radius, similarly to the Cas A CCO, which means that one cannot exclude the possibility that the X-ray emission comes from a fraction of the NS surface.\\

Overall, we do not see a compelling reason to prefer the carbon atmospheres over the hydrogen atmospheres. Spectral features would help to identify the correct atmosphere model. Unfortunately, the spectral features of the carbon atmosphere cannot be seen in the Cas A CCO spectrum because of the substantial interstellar absorption. Thus, there is currently no direct way to verify whether or not this NS is covered by a carbon atmosphere.\\

Note that in this work we did not consider a magnetic carbon atmosphere model. At the relatively low magnetic fields expected for CCOs (10$^{10}$ - 10$^{11}$ G), the cyclotron lines cannot change model continuum fluxes significantly (see Fig. 2 in \citealt{Suleimanov2010}). The effect of the magnetic field on the carbon atmosphere models can be more complicated due to the magnetic shift and broadening of the carbon spectral lines and photo-ionization edges, and an increasing line blanketing effect. 
This effect could change the model spectra quantitatively  and lead to some shift of the allowed region in the $M - R$ plane and the derived temperatures. 

\section{Summary}
There are no strong arguments to prefer either hydrogen or carbon atmospheres for the modeling of the X-ray spectrum of the Cas A CCO. 

For our data with negligible pile-up, the temperature change between 2006 and 2012 does not exceed the $2.5\sigma$ level applying either of those atmosphere models with various parameter configurations (emission area fixed, tied or free; $N_{\rm H}$ tied or free for the two epochs) at the default calibration.

If we allow a change of the emission area size, the best-fit temperature grows within its $1 \sigma$ uncertainty, while a temperature drop is found for tied or fixed emission area sizes. 
This temperature drop has a different significance depending on whether $N_{\rm H}$ is allowed to vary between the observations or not ($<1.0\sigma$ and $<2.5\sigma$ in the former and latter case, respectively).
If we consider the possible $\pm 10$\% uncertainty in the optical depth of the ACIS filter contamination, the significance of the temperature drop stays below $1.0\sigma$ for varying $N_{\rm H}$, but increases to $3.1\sigma$ if $N_{\rm H}$ is assumed to be the same in 2006 and 2012. From our CCO observations as well as observations of the Cas A SNR, we regard the former as the more realistic assumption.\\

For the case of different $N_{\rm H}$ and tied emission areas, for instance, we obtain $\triangle T = (-1.1\pm 1.9)\times 10^4$ K and the corresponding cooling rate is $-\dot{T} = 2000 \pm 3400$ K yr$^{-1}$ (errors at 90\% confidence level). 
The cooling rates obtained from fits to the results by HH10 ($-\dot{T}=7700\pm 1900$\,K\,yr$^{-1}$) or E+13 ($-\dot{T}=7700\pm 1300 $\,K\,yr$^{-1}$) overlap with our $-\dot{T}$ uncertainty at the $2\sigma$ level. 
Based on the broader range of models considered in this paper, the lower quality of the data used by HH10 and E+13 (uncertain CTI and pile-up correction), and the fact that E+13 only found a significant temperature drop for the Graded mode ACIS-S data but not, for instance, for the HRC observations, we believe that the previously reported rapid cooling of the Cas A CCO is likely a systematic artifact, and we cannot exclude the standard slow cooling for this NS.\\ 

In contrast to the temperature change, we see an apparent flux drop for \emph{all} considered atmosphere models. 
The decrease in the absorbed flux is $4$\%$-6$\% in the energy range $0.6-6$\,keV, and its significance is at the $\approx 2\sigma - 3\sigma$ level. The flux drop is most pronounced in the data for energies $\approx 1.4 - 1.8$\,keV. Although the model-fitted flux decrease remains below the $3\sigma$ significance level, the concerted behavior of the data points hints at a systematic effect in this energy range. Possible explanations are unaccounted  instrument calibration issues (e.g., individual ACIS contaminant components accumulate differently) and background/foreground uncertainties (e.g., small patches of Silicon-rich cold dust in the center of the SNR might have moved into the line of sight toward the CCO).\\

Overall, our results (2006-2012) are consistent with no temperature decline at all, or a smaller temperature decline than that reported for the data suffering from pile-up and acquired in Graded mode during the time interval 2000-2012.  
A longer time base of data with negligible pile-up and a better knowledge of the ACIS filter contamination changes are needed to assess any temperature or flux change with higher certainty. \\

{\it Facility:} \facility{Chandra}\\

\acknowledgments
We thank the staff of the Chandra X-ray Center (CXC) for their advice regarding the contamination of the ACIS detector. In particular, we would like to thank  A. Vikhlinin, C. Grant, J. McDowell and M. Nowak for helpful discussions and clarifications.
We are indebted to Keith Arnaud for his help in solving several issues with XSPEC.
We also thank Craig Heinke for helpful referee comments and suggestions.\\ 
 
The scientific results reported in this article are based on observations made by the Chandra X-ray Observatory.
Support for this work was provided by the National Aeronautics and Space Administration through Chandra Award Number G02-13083X issued by the Chandra X-ray Observatory Center, which is operated by the Smithsonian Astrophysical Observatory for and on behalf of the National Aeronautics Space Administration under contract NAS8-03060. Support for this work was also provided by the ACIS Instrument Team contract SV4-74018 issued by the Chandra X-ray Observatory Center and NASA grant NNX09AC84G.\\

V.S. was supported by the German Research Foundation (DFG) grant SFB/Transregio 7 "Gravitational Wave Astronomy", and the Russian Foundation for Basic Research (grant 12-02-97006-r-povolzhe-a).\\

This research has made use of SAOImage DS9, developed by
SAO; and SAO/NASA's Astrophysics Data System Bibliographic Services, and the CXC software CIAO and ChIPS.

\appendix
\section{Contamination of the ACIS chip}
\label{contamtest1}
A contaminant has been accumulating on the optical-blocking filters of the ACIS detectors. Therefore, the effective low-energy quantum efficiency of the detector is becoming lower over time. Details and reports about the effects of contaminant on the ACIS response can be found on the \chan webpages\footnote{For example at \texttt{cxc.harvard.edu/ciao/why/acisqecontam.html}}. 
There is a gradient in the amount of material on the filters with a higher thickness of the contaminant at the chip edges.
Within about 100 pixels of the outer edges of the  ACIS chips, the gradient is relatively steep\footnote{See, e.g., Fig.\,6 in\\ cxc.harvard.edu/ciao/download/papers/spatial\_contam\_effects.pdf}. As noted in Section~\ref{obsred}, the source centroid is at ${\rm CHIPY} \approx 50$ pixels in both of our observations. The instrument responses are different in 2006 and 2012 (see Figure~\ref{contameff}).\\

The contamination is studied and modeled in detail by the \chan X-ray Center.
Formally, a contamination transmission function is defined in the ACIS Spatial Contamination Effects memo\footnote{cxc.harvard.edu/ciao/download/papers/spatial\_contam\_effects.pdf} in the following way:
\begin{equation}
\epsilon (E,t,x,y) = \sum\limits_{j=1}^{M} F_j {\rm exp} \left[{-\kappa_j \sum_{i=1}^{N} \mu_{ij}(E) l_{ij} (t,x,y)}\right],
\label{trans}
\end{equation}
where the transmission efficiency, $\epsilon$, of $N$ contaminants, $i$, in $M$ partial covering regions, $j$, depends on 
the photon energy, $E$, the time $t$, the absorption coefficient per unit thickness, $\mu_{ij}(E)$, and the thicknesses of the contaminants, $l_{ij}$. $F_j$ and $\kappa_j$ are weights for partial filling factors of different contaminants. The thickness of the contaminant $i$ in the partial covering region $j$ is expressed as
\begin{equation}
 l_{ij} (t,x,y) = l_{0ij} (t-t_0) +  l_{1ij} (t-t_0) f_{ij}(x,y) 
\label{thickness}
\end{equation}
The function  $f_{ij}(x,y)$  defines the spatial distribution, and $t_0$ is the \chan reference time, MJD 50814.0. 
The product of the absorption coefficient and thicknesses yields the optical depth, $\tau$, which is measured and analyzed by the \chan calibration team.
In the course of the \chan data reduction, the parameters of the modeled contamination are taken into account by the CIAO software in form of the CALDB contamination file\footnote{The latest version is acisD1999-08-13contamN0007.fits}; for details on this file see the CXC Archive Interface Control Document, section 2.6 \footnote{See space.mit.edu/ASC/docs/ARD\_ICD/ARD\_ICD.ps.gz}.\\

Monitoring of the optical depth showed a potential slight change in the accumulation rate of the contaminant during 2012 (A. Vikhlinin, {\emph{pers. com.}})\footnote{See also space.mit.edu/$\sim$cgrant/qe/index.html}. At the same time, \citet{Helder2012} reported on increasing differences in flux measurements with XMM-$Newton$ and \chan in the monitoring campaign of SN 1987A. These recent developments have yet to be implemented in the contamination model, and the current CALDB contamination file (N0007) does not account for it. 
To assess the effect of an uncertain contamination, we change the contamination file for the 2012 spectrum extraction.  We define $\mu^{\rm new}_{ij}(E) = a \times \mu_{ij}(E)$, which modifies the optical depth, $\tau$, by an energy-independent factor. As a note of caution, we emphasize that this very simplistic modification does not account for other possible changes such as varying contributions of individual contaminant components which could cause energy-dependent changes of the contaminant optical depth.\\

\section{Carbon atmosphere models}
\label{carbmodels}
We developed new carbon atmosphere models\footnote{The carbon atmosphere models are now available in XSPEC: heasarc.gsfc.nasa.gov/xanadu/xspec/models/carbatm.html} for the spectral fits of the Cas A CCO.
The carbon model atmospheres were computed assuming hydrostatic and radiative
equilibria in a plane-parallel approximation. 
The basic assumptions of the stellar atmosphere modeling and all the used equations 
can be found in \citet{Mihalas1978}. The main input parameters are
the surface gravity
\begin{equation}
g=\frac{G M_{\rm NS}}{R^2_{\rm NS}}(1+z),
\label{eq:gdef}
\end{equation}
and the effective temperature $T_{\rm eff}$. 
The gravitational redshift $z$ on the star
surface is related to the NS mass $M_{\rm NS}$ and NS radius $R_{\rm NS}$ as follows:
\begin{equation}
1+z=\left( 1-\frac{2 G M_{\rm NS}}{c^{2} R_{\rm NS}} \right)^{-1/2} .
\label{eq:zdef}
\end{equation}

In our calculations, we assumed local thermodynamic equilibrium (LTE).
In addition, we accounted for the dissolving of bound levels and the 
pressure ionization effects 
using the occupation probability formalism \citep{Hummer1988}
as described by \citet{Hubeny1994}.
We took into account the coherent electron scattering as well as the free-free and
bound-free transitions for all carbon ions using the opacities from
\citet{Verner1996} and \citet{Verner1995}
\citep[see][]{Ibragimov2003}. The photoionization from the first 
excited levels of C{\sc{V}} and C{\sc{VI}} ions was also included. 
The spectral lines of these ions were taken into account from the
CHIANTI, Version 3.0, atomic database \citep{Dere1997}.

\begin{figure}[b]
\centering
{\includegraphics[width=85mm, clip]{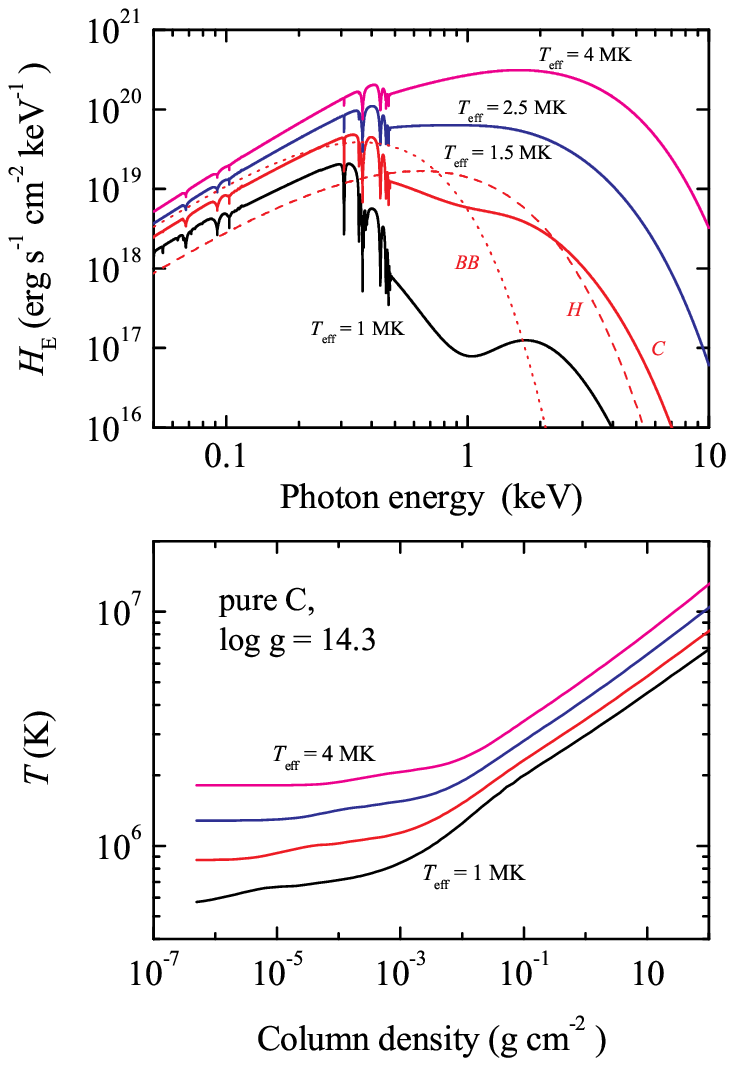}}
\caption{The emergent spectra of pure carbon atmospheres with a fixed
$\log g$ = 14.3 and a set of effective temperatures. For $T_{\rm eff}= 1.5$\,MK the blackbody spectrum (dotted curve) and the pure hydrogen model 
spectrum (dashed curve)
are also shown.}
\label{fig:af1}
\end{figure}

For computations, we used the numerical code ATLAS \citep{Kurucz1970,Kurucz1993} modified to deal with high temperatures \citep{Ibragimov2003,Suleimanov2007,Rauch2008}.
Using this code, we computed an extended grid of the carbon model atmospheres.
In this grid the models were computed for $\log g $ values from 13.70 to 14.90 
with a step of 0.15 and the effective temperatures from $T_{\rm eff} = 10^6$\,K to $4 \times 10^6$\,K with
a step size of $5 \times 10^4$\,K. 
The emergent spectra and temperature structures for a few 
effective temperatures and the fixed surface gravity are shown in 
Figure\,\ref{fig:af1}. 

The C{\sc{V}} and C{\sc{VI}} photoionization edges change the
spectral continuum dramatically in comparison with the blackbody spectrum or the spectrum of hydrogen model atmospheres \citep[see also][]{Ho2009}.
Overall, our assumptions are similar to those of HH09, e.g., regarding the opacities, and the two models are very similar as demonstrated in Figure~\ref{carbonmodel}.

The photon fluxes for all the models have been integrated in narrow (0.02\,keV) energy bins over an energy grid from 0.01 to 20.00\,keV. These photon flux spectra have been converted into a FITS-table that can be imported as an \texttt{atable} model component in XSPEC.
Our method of the carbon atmosphere modeling and the validity of the 
adopted assumptions will be discussed in a separate paper, Suleimanov et al. (\textsl{submitted to ApJS}).\\

\begin{figure}[t]
\centering
{\includegraphics[height=85mm, bb=14 80 480 700, clip, angle=90]{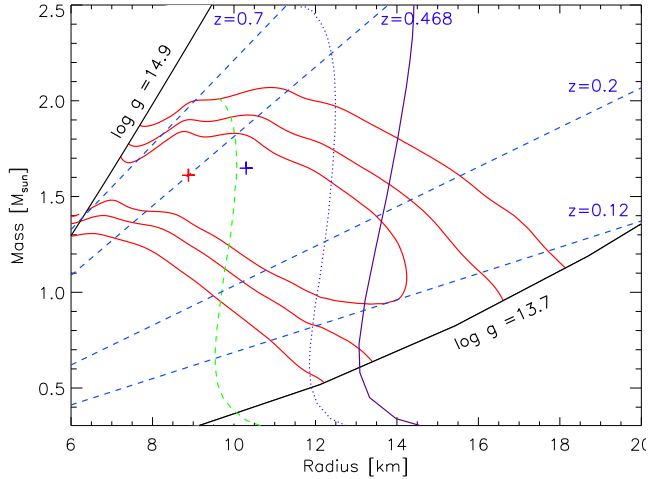}} 
\caption{Mass-radius confidence contours (68\,\%, 90\,\%, 99\,\%) for our carbon atmosphere model fit where $N_{\rm H}$ is allowed to be different in the two epochs. Note that these contours would become much broader if the additional uncertainties of the distance and of the emitting NS surface fraction were included. 
See the text for a detailed discussion.
For the case of a perfectly known distance and the assumption that the whole NS surface is emitting, our `best-fit model' ($\log g=14.6$, $z=0.468$ corresponding to $M_{\rm NS}=1.6$\,M$_{\odot}$ and  $R_{\rm NS}=8.9$\,km is marked with a red cross. 
The NS parameters used by HH10 ($M_{\rm NS}=1.648$\,M$_{\odot}$ and  $R_{\rm NS}=10.3$\,km) are marked with a blue cross. 
The two black lines indicate the range of the considered $\log g$ values, while the dashed blue lines indicate different gravitational redshift parameters.
The overplotted curves are the mass-radius relations for three possible EOSs according to \citet{Hebeler2013} (their Figures 11 and 12). The green dashed line corresponds to a soft EOS which decribes the minimal radius over the entire mass range, and the blue dotted line corresponds to an intermediate EOS. The purple solid line corresponds to the stiff EOS which follows closely the upper limit of the possible EOS range, and thus, indicates the largest possible NS radii.
\label{carbonMassRadius}}
\end{figure}

For our XSPEC fits, we probed a range of the gravitational redshift, $z$, adapted to the $\log g $ values in order to represent reasonable NS masses and radii. The $z$ sampling in our model grid was $\triangle z=0.006$.
Formally, the model having the lowest $\chi^2_{\nu}$ fit in the ($\log g$, $z$) grid is the `best-fit model', and confidence levels can be derived in the ($\log g $, $z$) space or the corresponding ($M_{\rm NS}$, $R_{\rm NS}$) space. 
Assuming that the whole NS surface is emitting in X-rays at a fixed distance $d=3.4$\,kpc, we obtained such `best-fit model' and its confidence contours. These contours are shown in Figure~\ref{carbonMassRadius} for the case of different $N_{\rm H}$ in the two epochs. They cover a large region in the mass-radius space. 
If we allowed the distance or the emitting area, hence the normalization, to be fit parameters\footnote{Note that we tied the normalizations in the two epochs to be the same, though.}, we were not able to produce reasonable confidence contours. 
This is due to the fact that the model parameters are highly correlated (e.g., the temperature with the gravitational redshift) and the spread in the resulting $\chi^2_{\nu}$ values is very small, even for very different ($\log g $, $z$) models.
Therefore, we want to emphasize that the confidence contours in Figure~\ref{carbonMassRadius} cover in fact an unrealistically small area since, first, there is a distance uncertainty and, second, it also remains to be proven that the temperature of the entire NS surface is uniform. 
Figure~\ref{carbonMassRadius} provides, however, a good representation of the considered ($\log g $, $z$) or corresponding ($M_{\rm NS}$, $R_{\rm NS}$) space as well as the location of the models discussed below. 
While we cannot derive useful constraints on mass and radius, thus on the NS equation of state (EOS), we show for completeness in Figure~\ref{carbonMassRadius} three representative EOSs provided by \citet{Hebeler2013}. 
These EOSs illustrate the center and the extremes of the allowed EOS region.  Our best-fit mass and radius are outside that allowed region due to a small radius; however, our 68\% confidence contour encompasses the whole radius range of these three representative EOS. Our 68\% confidence contours in mass-radius overlap with those previosly calculated for Cas A (e.g., \citealt{Yakovlev2011}, their Figure 1). The orientations of the confidence contours in the two works are slightly different due to the different number of free fit parameters in the models.  \\

Here, we want to illustrate the effect of the gravitational redshift on the temperature estimates, in particular the substantial uncertainty of the model parameters {{at the NS surface}} due to the unknown gravitational redshift. 
Our best-fit model (fixed distance, whole NS surface is emitting in X-rays) has the grid parameters $\log g=14.6$, $z=0.468$ which corresponds to $M_{\rm NS}=1.612$\,M$_{\odot}$ and  $R_{\rm NS}=8.881$\,km. 
For this model fit, we derive the effective temperatures $T_{4,2006}=220.2^{+1.3}_{-1.4}$ and $T_{4,2012}=219.1 \pm {1.5} $ (fit with $N_{\rm H}$ different in the two epochs; the errors indicate a 90\% confidence level).
Note that these values are the effective temperatures {\emph{at the NS surface}}. Any temperature constraint on the spectrum obtained by a distance observer, however, actually corresponds to $T^{\infty}_{\rm eff}=T_{\rm eff} (1+z)^{-1}$. This correlation between spectral model parameters\footnote{Because the normalization is defined as $R^2_{\rm NS}/ d^2$, and $R^{\infty}_{\rm NS}=R_{\rm NS} (1+z)$, the normalization is affected as well.} and the gravitational redshift is the reason why it is so difficult to constrain mass and radius from such atmosphere fits. 

The above temperatures in the case of ($\log g=14.6$, $z=0.468$), for example, correspond to $T^{\infty}_{4, 2006}=150.0^{+0.9}_{-1.0} $ and $T^{\infty}_{4,2012}=149.3 \pm {1.0} $. If we now consider a different case, e.g., $\log g=14.45$, $z=0.375$ ($M_{\rm NS}=1.647$\,M$_{\odot}$ and  $R_{\rm NS}=10.328$\,km), we obtain from a fit to the same model setup  $T_{4,2006}=201.1 \pm 1.2$, and $T_{4,2012}=200.1^{+1.3}_{-1.4}$, which correspond to $T^{\infty}_{4,2006}=146.3 \pm 0.9 $ and $T^{\infty}_{4,2012}=145.5 \pm 0.7$. 
The temperatures at infinity for different $M_{\rm NS}$, $R_{\rm NS}$ are much closer to each other than the temperatures at the NS surface. 
Furthermore, the temperature {\emph{difference}} between the epochs is the same at infinity taking rounding errors into account.
From this exercise, one concludes that it does not matter quantitatively which ($\log g$, $z$) pair one chooses to investigate the temperature evolution -- as long as this pair is located within our $1\sigma$ mass-radius contour.
In this paper we give all our fit results for  ($\log g=14.45$, $z=0.375$). This allows for an easy comparison with the results from HH10 who used ($\log g=14.4534$, $z=0.377$).\\

\section{Dust scattering}
\label{appdust}
In addition to the absorption by the interstellar gas,  X-ray 
photons can be absorbed and scattered by the interstellar dust. The dust
absorption is included in modern interstellar extinction models (e.g, 
`\texttt{tbabs}' in XSPEC), but the dust scattering is not. To take it into
account in the spectral analyis of a point X-ray source, the 
absorbed model source spectrum 
should be additionally multiplied by the scattering attenuation factor
$D_s(E) = \exp[-\tau_s(E)]$,
where $\tau_s(E)$ is the optical depth with respect to the dust scattering, and $E$ is the photon energy. In the Rayleigh-Gans approximation the scattering cross section is proportional to $E^{-2}$, and the attenuation
factor can be expressed as
\begin{equation}
D_s(E) = \exp[-S N_{\rm H,22} E^{-2}],
\label{modelattenuation}
\end{equation}
where $E$ is in units of keV, $N_{\rm H,22}$ is the hydrogen column density
in units of $10^{22}$ cm$^{-2}$, and $S = \tau_s(E=1\, {\rm keV})/N_{\rm H,22}$ is a 
factor proportional
to the average dust-to-gas ratio along the line of sight to the source
(see, e.g., Predehl \& Schmitt 1995). The $S$ factor has been measured for a number of bright X-ray sources based on observations of their
dust scattering halos. In particular, Predehl \& Schmitt (1995) found
a mean value,  $S=0.49$, from {\sl ROSAT} observations. 
The $S$ factor, however, 
shows a large scatter for different sources and different halo studies
(see the Appendix of \citet{Misanovic2011} for more details and 
references), which is not surprising given the possible different dust compositions and grain sizes, and different dust-to-gas ratios in different
parts of the ISM, as well as the simplified 
scattering model adopted. Including the dust scattering attenuation in
spectral fits not only reduces the inferred $N_{\rm H}$ value, but it also
might allow one to measure $S$, because the energy dependence of 
$D_s(E)$ is
different from that of the `true absorption' attenuation, $D_a(E)=\exp[-N_{\rm H} \sigma_{a,{\rm eff}}(E)]$, where $\sigma_{a,{\rm eff}}(E)$ is the effective
absorption cross section. Moreover, since
 the change of the extinction model may change the inferred values of
other fitting parameters, it seems important to allow for the dust scattering in spectral fits.
 However, unless the number of detected source counts
is very large, especially at higher energies, where the dust scattering
may dominate the true absorption, 
the $S$ parameter remains 
poorly constrained because of the strong (anti)correlation between $S$ and $N_{\rm H}$.
Even at a fixed $S$ value the inclusion of the dust scattering model 
just reduces the inferred $N_{\rm H}$ (the stronger the larger the chosen $S$ is),  without a substantial change of other
fitting parameters.
Since the effective absorption cross section at X-ray energies
 can be crudely approximated as
$\sigma_{\rm a, eff}(E) \sim 2\times 10^{-22} E^{-8/3}$ cm$^2$ \citep[page 196]{Allen1977}, the 
 hydrogen column
density estimated with allowance for the dust scattering is
$N_{\rm H}^{\rm dust} \sim  (1 + 0.5 S\bar{E}^{2/3})^{-1} N_{\rm H}$ (e.g., \citealt{Morrison1983}), where
$\bar{E}$ is a characteristic photon energy in the source spectrum, and $N_{\rm H}$ is estimated without including the dust scattering. In particular,
for $\bar{E}\sim 2$ keV and $S\sim 0.5$, we obtain 
$N_{\rm H}^{\rm dust} \sim 0.7 N_{\rm H}$.
 Because the account for the dust scattering just leads to such an ``$N_{\rm H}$ renormalization'' (uncertain because the $S$ value is not known a priori), the allowance for the dust scattering is not expected to
change the intrinsic source model spectrum in most cases.

HH09, HH10, \citet{Shternin2011}, and E+13
included a dust scattering model in their Cas A CCO X-ray spectral fits, 
referring 
to \citet{Predehl2003}, 
who used Equation (\ref{modelattenuation}) with $S=0.49$.
To check the effect of dust scattering on the parameters of the
Cas A CCO fits, we used the XSPEC `{\em dust}' model
\begin{equation}
D_{s,{\rm XSPEC}}(E) = 1 - 
p_1\left(\frac{1}{E^2} - \frac{1}{p_2^2}\right),
\label{xspecdust}
\end{equation}
where $p_1$ and $p_2$ are the model parameters (K. Arnaud, {\emph{priv.\ comm.}}).
The last term in Equation (\ref{xspecdust}), $p_1/p_2^2$, 
takes into account 
the contribution of the dust scattering halo in the point source aperture,
negligible at the {\sl Chandra} angular resolution. Without this term,
Equation (\ref{xspecdust}) is equivalent to 
Equation (\ref{modelattenuation}) in the optically thin approximation,
i.e., $p_1 = S N_{\rm H,22}$. 
Fitting the Cas A CCO's 2006 spectrum  with the carbon atmosphere model
 at $p_1=0.49 N_{\rm H,22}$
(and $p_2\to \infty$), we can reproduce the $N_{\rm H}$ obtained by \citet{Ho2009} -- our $N_{\rm H,22}=1.55 \pm 0.04$ 
is consistent with the HH09 value, $N_{\rm H,22}=1.65 \pm 0.05$,
at the $2\sigma$ level.
Furthermore, we found that the additional {\em dust} model component does not improve the fit: $\chi^2_{\nu}=1.02$ (with dust scattering) versus $\chi^2_{\nu}=1.00$ (without dust scattering) for the fit of the 2006 data.
Moreover, we found that the temperatures and their difference are not significantly affected by the 
inclusion of the dust scattering model, as expected. In the case of variable  $N_{\rm H}$ in the two epochs, for example, we obtain $T_{4,2006}=201.3 \pm 1.2$,
and $T_{4,2012}=200.3^{+1.2}_{-1.3}$
(with dust scattering) in comparison to $T_{4,2006}=201.1 \pm 1.2$,
and $T_{4,2012}=200.1^{+1.3}_{-1.4}$
(without dust scattering).
Given the uncertainties and possible changes in the dust properties in the dynamic Cas A SNR, and the lack of effect on the inferred temperature differences,
we believe there is no need to include the dust scattering 
in the investigation of temperature variations.

\bibliographystyle{apj}
\bibliography{Casa}

\end{document}